\documentclass[prb,nofootinbib,twocolumn,superscriptaddress]{revtex4} 
\usepackage{graphicx}% Include figure files
\usepackage{dcolumn}% Align table columns on decimal point
\usepackage{bm}% bold math 
\usepackage{threeparttable}
\usepackage{times}
\usepackage{mathptmx}
\usepackage{lscape}
\usepackage{natbib}
\usepackage{amsmath}
\usepackage{amssymb}
\usepackage{braket}
\usepackage{comment}
\usepackage{color}

\def\degree{\kern-.2em\r{}\kern-.3em}

\begin{document}

%\preprint{APS/123-QED} 

\title{  Does Stochastic Disorder Conform to Configurational Disorder?  }

\author{Shouno Ohta}
\affiliation{
Department of Materials Science and Engineering,  Kyoto University, Sakyo, Kyoto 606-8501, Japan\\
}%

\author{Koretaka Yuge}
\affiliation{
Department of Materials Science and Engineering,  Kyoto University, Sakyo, Kyoto 606-8501, Japan\\
}%

\begin{abstract}
{ In alloy thermodynamics, stochastically disordered state (SDS), where each lattice point is stochastically occupied by constituents according to given composition, is typically referred to investigating physical properties for homogeneously substitutional state: The so-called special quasirandom structure (SQS) of a \textit{single} microscopic structure, that mimics multisite correlation function for SDS, is amply, widely used for bulk, surface, interface and nano-cluster properties. Despite the widely-used concept for SDS, it has not been clear whether the SDS should conform to configurationally disordered state (CDS) for discrete system, i.e., average over all possible configuration. Here we quantitatively discuss the difference between SDS and CDS for multisite correlation, and show the condition where SDS conforms to CDS. The results show that when practical system size contains below $\sim 100,000$ atoms, differences in multisite correlation between SDS and CDS remains few percent depending on the geometry of lattice as well as of figure, indicating that SQS for subsystem of surface and interface, and for isolated clusters, should be carefully applied to investigate high-temperature properties as CDS. 
  }
\end{abstract}

%\pacs{81.90.+c \sep 61.05.-a \sep 05.20.Gg \sep 05.10.-a \sep 02.30.Zz }

\maketitle

\section{Introduction}
%\begin{itemize}
%\item
In alloy thermodynamics, expectation value of dynamical variables (e.g., internal energy, vibrational free energy and elastic modulus) in equilibrium state at given composition is typically given by the canonical average,
\begin{eqnarray}
\Braket{C} = Z^{-1}\sum_i C_i \exp{\left(-\beta E_i\right)},
\end{eqnarray}
where summation is taken over possible microscopic states on configuration space, and 
$Z$ denotes partition function. 
%\item
Since the number of microscopic states exponentially increases with increase of system size, exact estimation of macroscopic property is typically far from practical.
Therefore, variety of theoretical approaches has been developed to effectively sample important states for macroscopic property, including Metropolis algorithm, entropic sampling and Wang-Landau sampling.\cite{mc1,mc2,mc3,wl} 

%\item
An alternative approach has been proposed to investigate the properties at homogeneously substitutional state (i.e., there is no short- or long-range order), where corresponding disordered state is mimicked by a single microscopic state, spesial quasirandom structure (SQS).\cite{sqs} 
 %\item
SQS has been widely used for investigating bulk, surface, interface and nano-cluster properties refered as \textit{homogeneous} substitutional state, corresponding to configurationally disordered state (CDS), while the concept of SQS is  based on stochastically  disordered state (SDS), i.e., its multisite correlation is determined where each lattice point is occupied according to given probability for composition. 
%\item
Although the difference between CDS and SDS is expected to be negligible at thermodynamic limit of system size $N\to\infty$, it is unclear whether application of SQS to mimic homogeneous state for practical finite-size system, or locally-equilibrium sub-system including surface and interface, is justified: It has not been clarified to what extent SQS should conform to CDS in terms of the system size and of geometry for underlying lattice and considered figure. 
%\item
Here we theoretically demonstrate that for classical discrete system on periodic lattice under constant composition, how SQS (or SDS) exactly differs from CDS, and the condition where SDS conforms to CDS. 
The details are shown below.
%\end{itemize}

\section{Derivation and Discussions}
%\begin{itemize}
%\item
\subsection{SDS and CDS: Differences at thermodynamic limit}
\subsubsection{Binary system with conventional basis functions}
We first individually define the correlation function for SDS (or SQS) and CDS for A$\left(1-x\right)$B$_x$ binary system. 
We employ generalized Ising model\cite{ce} (GIM) to quantitatively describe all possible atomic configuration, where its basis function for $m$-body figure $c$ is given by the linear average of spin product contains in the figure:
\begin{eqnarray}
\phi_c = \Braket{\prod_{i\in c}\sigma_i}_{\textrm{lattice}},
\end{eqnarray}
where $\sigma_i = +1$ (-1) when site $i$ is occupied by A (B) atom. With this definition, on-site correlation (i.e., figure has a single lattice point) can be relates to the composition $x$,  $\phi_o = 1-2x$. 
%\item
Therefore, multisite correlation function of $m$-body figure for SQS can be simply given by 
\begin{eqnarray}
\label{eq:sqs}
\phi^{\left(m\right)}_{\textrm{SQS}} = \left(1-2x\right)^m,
\end{eqnarray}
corresponding to independent, random occupation of each lattice point according to given composition, $x$. 
%\item
Meanwhile, multisite correlation function for CDS is not trivially estimated, since it is a linearly-averaged value over all microscopic states on configuration space under given system size, $N$. 
Our recent study provides exact formulation of multisite correlation function for CDS, given by\cite{ex}
\begin{widetext}
\begin{eqnarray}
	\label{eq::average_binary:1}
	\begin{split}
		\Braket{\xi} 
		&=
			\dfrac{1}{{}_N {\rm C}_{x N}}
			\displaystyle\sum_{k=1}^{m}
			{}_m {\rm C}_k
			\cdot
			{}_{N-m} {\rm C}_{x N -k}
			\cdot
			\left(-1\right)^k
		\\[5pt]
		&=
			\dfrac{\left(1 - x\right) N ! \cdot x N !}{N !}
			\displaystyle\sum_{k=1}^{m}
			\dfrac{\left( N - m \right) ! \cdot N^m \cdot {}_m {\rm C}_k \left(-x\right)^k \left( 1-x \right)^{ m-k }}
			{\left(\left(1-x\right) N - \left( m-k \right)\right) ! \cdot \left\{ \left( 1- x \right) N \right\}^{\left( m-k \right)} \cdot \left(x N - k\right) ! \cdot \left( x N \right)^k}
	\end{split}
\end{eqnarray}
\end{widetext}
At thermodynamic limit of $N\to\infty$, terms included in the summation of Eq.~(\ref{eq::average_binary:1}) can be given by
\begin{widetext}
\begin{eqnarray}
	&\left( N - m \right) ! \cdot N^m 
	=
	N ! \cdot \left( 1 \cdot \dfrac{N}{N-1} \cdots \dfrac{N}{N-\left( m-1 \right)}\right)
	\underset{N\rightarrow\infty}{\longrightarrow} N!&\\[5pt]
	&\begin{split}
		&\left(\left(1-x\right) N - \left( m-k \right)\right) ! \cdot \left\{ \left( 1- x \right) N \right\}^{\left( m-k \right)}\\
		&\hspace{70pt} =
		\left(1-x\right) N ! \cdot \left( 1 \cdot \dfrac{\left(1-x\right) N}{\left(1-x\right) N - 1} \cdots  \dfrac{\left(1-x\right) N}{\left(1-x\right) N - \left( m-k-1 \right)} \right)
		\underset{N\rightarrow\infty}{\longrightarrow} \left(1-x\right) N !
	\end{split}&\\[5pt]
	&\left(x N - k \right) ! \cdot \left( x N \right)^{k}
	=
	x N ! \cdot \left( 1 \cdot \dfrac{x N}{x N - 1} \cdots  \dfrac{x N}{x N - \left( k-1 \right)} \right)
	\underset{N\rightarrow\infty}{\longrightarrow} x N !, &
\end{eqnarray}
\end{widetext}
which directly leads to
\begin{widetext}
\begin{eqnarray}
	\label{eq::average_binary:2}
	\begin{split}
		\Braket{\xi} 
		\underset{N\rightarrow\infty}{\longrightarrow}
		\dfrac{\left(1 - x\right) N ! \cdot x N !}{N !}
		\displaystyle\sum_{k=1}^{m}
		\dfrac{N !}{\left(1 - x\right) N ! \cdot x N !}
		\cdot
		{}_m {\rm C}_k
		\left(-x\right)^k
		\left( 1-x \right)^{ m-k }
		%\\[5pt]
		=
		\displaystyle\sum_{k=1}^{m}
		{}_m {\rm C}_k
		\left(-x\right)^k
		\left( 1-x \right)^{ m-k }
		~=~
		\left( 1-2x \right)^m .
	\end{split}
\end{eqnarray}
\end{widetext}
Therefore, we can clearly see that SDS of Eq.~(\ref{eq:sqs}) conforms to CDS at thermodynamic limit for binary system at any order of multisite correlations at any composition, $x$.

\subsubsection{Generalization to multicomponent system with any basis functions}
Now let us generalize the differences between SDS and CDS on binary system, to multicomponent systems with any choice of orthonormal basis functions. 
Here we consider $r$-component system with individual composition of $x_0,\cdots,x_{r-1}$ and number of atoms, $N$. 
Orthonormal basis functions at a single lattice point is given by $\theta^{\left(0\right)},\cdots,\theta^{\left(r-1\right)}$ where we always choose $\theta^{\left(0\right)}=1$ without lack of generality. 
Value of spin variables for individual components $X_0,\cdots,X_{r-1}$ are given by $s_0,\cdots,s_{r-1}$. 
For instance, when a given lattice point $p$ is occupied by $X_k$, the value for corresponding chosen basis function is $\theta^{\left(l\right)}\left(\sigma_p\right)=\theta^{\left(l\right)}\left(s_k\right)$.
Basis functions for whole lattice points can therefore be obtained by taking tensor product of vector space spanned by basis functions on individual lattice point.
Then, correlation function (basis function) for given combination of basis functions on all lattice points at given atomic configuration $\vec{\sigma}=\left\{\sigma_1,\cdots,\sigma_N\right\}$ can be 
given by
\begin{eqnarray}
	\label{eq::average_multi:XI_DEFINITION}
	\xi_{\left\{l\right\}} \left( \overrightarrow{\sigma} \right)
	=
		\theta^{\left({l_1}\right)}\left( \sigma_1 \right)
		\cdots
		\theta^{\left({l_N}\right)}\left( \sigma_N \right) ,
\end{eqnarray}
where number of $\theta^{\left(l\right)}$ included in the r.h.s. of Eq.~(\ref{eq::average_multi:XI_DEFINITION}) is defined as $m^{\left(l\right)}$.
(e.g., correlation function with number of $\theta^{\left(1\right)}$ to be $m$ is called $m$-body correlation for binary system).
For generalized Ising systems, correlation function is generally defined as linear average of the above equation over all symmetry-equivalent figures on latice, i.e., 
$\xi \left( \overrightarrow{\sigma} \right) = \Braket{\xi_{\left\{l\right\}}}_{\left\{l\right\}} \left(\overrightarrow{\sigma}\right)$ for given atomic configuration.
1-order moment is given by taking average of $\xi \left( \overrightarrow{\sigma} \right)$ over all possible configurations, $\Braket{\xi} = \Braket{\xi \left( \overrightarrow{\sigma} \right)}_{\overrightarrow{\sigma}}$. 
For our purpose to provide exact formulation of 1-order moment for CDS, this can be written by
\begin{widetext}
\begin{eqnarray}
	\label{eq::average:XI_MOM1}
	\Braket{\xi}
	=
		\dfrac{\left(x_0 N\right) ! \cdots \left(x_{r-1} N\right) !}{N !}
		\displaystyle \sum_{\star 1}
		\left[
		\displaystyle\prod_{l=0}^{r-1}
		\left\{
		\dfrac{m^{\left(l\right) } !}{m^{\left(l\right)}_0 ! \cdots m^{\left(l\right)}_{r-1} !}
		\left\{\theta^{\left({l}\right)}\left( s_0 \right)\right\}^{m^{\left(l\right)}_0}
		\cdots
		\left\{\theta^{\left({l}\right)}\left( s_{r-1}\right)\right\}^{m^{\left(l\right)}_{r-1}}
		\right\}
		\right],
\end{eqnarray}
\end{widetext}
where $m^{\left(l\right)}_{k}$ denotes number of lattice points occupied by component $X_k$ with basis function $\theta^{\left(l\right)}$, and summation of $\star 1$ is taken 
for all possible combination of $\left\{m_l^{\left(k\right)}\right\}$ satisfying the following condition:
\begin{eqnarray}
	\star 1 ~:~ \displaystyle \sum_{l=0}^{r-1} m^{\left(l\right)}_k = x_k N,~\displaystyle\sum_{k=0}^{r-1} m_k^{\left(l\right)} = m^{\left(l\right)},~m^{\left(l\right)}_k \geq 0.
\end{eqnarray}
Eq.~(\ref{eq::average:XI_MOM1}) can be further rewritten by explicitly considering the number of ways for lattice points with $\theta^{\left(0\right)}=1$ and other lattice points with other basis functions of non-unity, namely
\begin{widetext}
\begin{eqnarray}
	\label{eq::average:XI_MOM2}
	\begin{split}
			&\displaystyle\sum_{\star 2}
			\left[
			\dfrac{\left(x_0 N\right) ! \cdots \left(x_{r-1} N\right) !}{N !}
			\dfrac{\left(N - \sum_{l=1}^{r-1} m^{\left(l\right)}\right) !}{\left( x_0 N - \sum_{l=1}^{r-1} m^{\left(l\right)}_0 \right) ! \cdots \left( x_{r-1} N - \sum_{l=1}^{r-1} m^{\left(l\right)}_{r-1} \right) !}
			\right.
		\\[5pt]
		&\hspace{150pt}
			\left.
			\cdot
			\displaystyle\prod_{l=1}^{r-1}
			\left\{
			\dfrac{m^{\left(l\right)} !}{m^{\left(l\right)}_0 ! \cdots m^{\left(l\right)}_{r-1} !}
			\left(
			\displaystyle\prod_{k=0}^{r-1}
			\left\{\phi^{\left(l\right)} \left(s_k\right) \right\}^{m^{\left(l\right)}_k}
			\right)
			\right\}
			\right],
	\end{split}
\end{eqnarray}
\end{widetext}
where summation of $\star 2$ is taken for $m^{\left(k\right)}_l$ satisfying the followings:
\begin{eqnarray}
\star 2 ~:~ \displaystyle \sum_{l=1}^{r-1} m^{\left(l\right)}_k \leq m_k,~\displaystyle\sum_{k=0}^{r-1} m_k^{\left(l\right)} = m^{\left(l\right)},~m^{\left(l\right)}_k \geq 0.
\end{eqnarray}
When we define that 1-order moment of point correlation function $y^{\left(l'\right)}$ as a product of a single $\theta^{\left(l'\right)}$ and $N-1$ functions of $\theta^{\left(0\right)}$, 
we can obtain $y^{\left(l'\right)}$ by simply substituting $m^{\left(l'\right)} = 1,~m^{\left(l\right)} = 0 \left( l \ne 0,l'\right)$ into Eq.~(\ref{eq::average:XI_MOM2}): 
\begin{widetext}
\begin{eqnarray}
	\label{eq::average_multi:XI_1STforPOINT1}
	\begin{split}
		y^{\left(l'\right)} 
		&=
			\dfrac{\left(x_0 N\right) ! \cdots \left(x_{r-1} N\right) !}{N !}
			\displaystyle\sum_{\star 2}
			\left[
			\dfrac{\left(N-1\right) !}{\left(x_0 N-m^{\left({l'}\right)}_0\right) !
			\cdots
			\left(x_{r-1} N- m^{\left({l'}\right)}_{r-1}\right) !}
			\right.
		\\
		&\left.\hspace{150pt}
			\cdot
			\dfrac{1}{m^{\left({l'}\right)}_0 ! \cdots m^{\left({l'}\right)}_{r-1} !} \left\{\theta^{\left({l'}\right)}\left( s_0 \right)\right\}^{m^{\left({l'}\right)}_0}
			\cdots
			\left\{\theta^{\left({l'}\right)}\left( s_{r-1} \right)\right\}^{m^{\left({l'}\right)}_{r-1}}
			\right]
		\\[20pt]
		&=
			\dfrac{\left(x_0 N\right) ! \cdots \left(x_{r-1} N\right) !}{N !} 
			\displaystyle\sum_{k=0}^{r-1}
			\left[
			\dfrac{\left(N-1\right) !}{\left(x_1 N\right) ! \cdots \left(x_{k} N - 1\right) ! \cdots \left( x_{r-1} N\right) !}
			\left\{ \theta^{\left(l'\right)} \left( s_k \right) \right\} \right]
		~=~
			\displaystyle\sum_{k=0}^{r-1}
			\left\{ \theta^{\left(l'\right)} \left(s_k\right) \right\} \cdot x_k.
	\end{split}
\end{eqnarray}
\end{widetext}
Then, we consider more general case of 1-order moment of $m^{\left(1\right)} + \cdots + m^{\left(r-1\right)} $-body correlation including $\left\{m^{\left(l\right)}\right\}$ basis functions of $\left\{\theta^{\left(l\right)}\right\}$, which is given by
\begin{widetext}
\begin{eqnarray}
	\label{eq::average_multi:XI_1STforMULTIBODY1}
	\begin{split}
		\Braket{\xi}
		&=
			\dfrac{\left(x_0 N\right) ! \cdots \left(x_{r-1} N\right) !}{N !}  
			\displaystyle\sum_{\star 2} 
			\left[ 
			\dfrac{\left( N-\sum_{l=1}^{r-1} m^{\left(l\right)} \right) !}
			{\left( x_0 N-\sum_{l=1}^{r-1}m^{\left(l\right)}_0 \right) ! \cdots \left( x_{r-1} N-\sum_{l=1}^{r-1}m^{\left(l\right)}_{r-1} \right) !}
			\right.
		\\
		&\left.\hspace{120pt}
			\cdot
			\displaystyle\prod_{l=1}^{r-1}
			\left\{
			\dfrac{m^{\left(l\right) } !}{m^{\left(l\right)}_0 ! \cdots m^{\left(l\right)}_{r-1} !}
			\left\{\theta^{\left({l}\right)}\left( s_0 \right)\right\}^{m^{\left(l\right)}_0}
			\cdots
			\left\{\theta^{\left({l}\right)}\left( s_{r-1} \right)\right\}^{m^{\left(l\right)}_{r-1}} 
			\right\}
			\right]
		\\[30pt]
		&=
			\dfrac{\left(x_0 N\right) ! \cdots \left(x_{r-1} N\right) !}{N !}
		\\[5pt]
		&\hspace{0pt}
			\cdot
			\displaystyle\sum_{\star 2}
			\left[ 
			\dfrac{
			\left( N-\sum_{l=1}^{r-1} m^{\left(l\right)} \right) ! \cdot {N}^{\sum_{l=1}^{r-1} m^{\left(l\right)}}
			}
			{
			\left( x_0 N-\sum_{l=1}^{r-1}m^{\left(l\right)}_0 \right) ! \cdot \left(x_{0} N\right)^{\sum_{l=1}^{r-1}m^{\left(l\right)}_0}
			\cdots
			\left( x_{r-1} N-\sum_{l=1}^{r-1}m^{\left(l\right)}_{r-1} \right) ! \cdot \left(x_{r-1} N\right)^{\sum_{l=1}^{r-1}m^{\left(l\right)}_{r-1}}}
		\right.\\[5pt] 
		&\left.\hspace{40pt}
			\cdot
			\left( \prod_{k=0}^{r-1} {x_k}^{\sum_{l=1}^{r-1} m^{\left(l\right)}_k} \right)
			\cdot
			\displaystyle\prod_{l=1}^{r-1}
			\left\{
			\displaystyle
			\dfrac{m^{\left(l\right) } !}{m^{\left(l\right)}_0 ! \cdots m^{\left(l\right)}_{r-1} !} 
			\left\{\theta^{\left({l}\right)}\left( s_0 \right)\right\}^{m^{\left(l\right)}_0}
			\cdots
			\left\{\theta^{\left({l}\right)}\left( s_{r-1} \right)\right\}^{m^{\left(l\right)}_{r-1}} \right\}
			\right].
	\end{split}
\end{eqnarray}
\end{widetext}
The last equation, we artificially introduce additional terms in analogy to taking thermodynamic limit for binary system, i.e., Eqs.~(\ref{eq::average_binary:1})-(\ref{eq::average_binary:2}). 
Therefore, in a similar fashion to derivation in binary system, we can see the thermodynamic limit for terms in Eq.~(\ref{eq::average_multi:XI_1STforMULTIBODY1}):
\begin{widetext}
\begin{eqnarray}
	&\begin{split}
		\left( N-\sum_{l=1}^{r-1} m^{\left(l\right)} \right) ! \cdot {N}^{\sum_{l=1}^{r-1} m^{\left(l\right)}}
		=
		N ! \cdot \left( 1 \cdot \dfrac{N}{ N-1 } \cdots \dfrac{N}{N- \left( \sum_{l=1}^{r-1} m^{\left(l\right)}-1 \right)} \right)
		\underset{N\rightarrow\infty}{\longrightarrow} N !
	\end{split}&\\
	&\begin{split}
		&\left( x_k N-\sum_{l=1}^{r-1} m^{\left(l\right)}_k \right) ! \cdot {x_k N}^{\sum_{l=1}^{r-1} m^{\left(l\right)}_k}\\
		&\hspace{70pt}=
		\left(x_k N\right) ! \cdot \left( 1 \cdot \dfrac{x_k N}{\left( x_k N-1 \right) !} \cdots \dfrac{x_k N}{x_k N- \left( \sum_{l=1}^{r-1} m^{\left(l\right)}_k-1 \right)} \right)
		\underset{N\rightarrow\infty}{\longrightarrow} x_k N !.
	\end{split}&
\end{eqnarray}
\end{widetext}
Using these results, we can derive that multisite correlations for multicomponent system with any basis functions in CDS conform to the product of 1-order moment for point correlation (i.e., SDS) at thermodynamic limit, as followings:
\begin{widetext}
\begin{eqnarray}
\label{eq::average_multi:XI_1STforMULTIBODY2}
	\begin{split}
	\Braket{\xi}
	&\underset{N\rightarrow\infty}{\longrightarrow}
		\dfrac{\left(x_0 N\right) ! \cdots \left(x_{r-1} N\right) !}{N !}
		\cdot
		\displaystyle\sum_{\star 2}
		\left[\dfrac{N !}{\left(x_0 N\right) ! \cdots \left(x_{r-1} N\right) !} 
	\right.\\[5pt] 
	&\left.\hspace{40pt}
		\cdot
		\left( \prod_{k=0}^{r-1} {x_k}^{\sum_{l=1}^{r-1} m^{\left(l\right)}_k} \right)
		\cdot
		\displaystyle\prod_{l=1}^{r-1}
		\left\{
		\dfrac{m^{\left(l\right) } !}{m^{\left(l\right)}_0 ! \cdots m^{\left(l\right)}_{r-1} !}
		\left\{\theta^{\left({l}\right)}\left( s_0 \right)\right\}^{m^{\left(l\right)}_0}
		\cdots
		\left\{\theta^{\left({l}\right)}\left( s_{r-1} \right)\right\}^{m^{\left(l\right)}_{r-1}} 
		\right\}
		\right]
	\\[40pt]
	&=
		\displaystyle\sum_{\star 2}
		\left[
		\left( \prod_{k=0}^{r-1} {x_k}^{\sum_{l=1}^{r-1}m^{\left(l\right)}_k} \right)
		\cdot
		\displaystyle\prod_{l=1}^{r-1}
		\left\{
		\dfrac{m^{\left(l\right) } !}{m^{\left(l\right)}_0 ! \cdots m^{\left(l\right)}_{r-1} !} 
		\left\{\theta^{\left({l}\right)}\left( s_0 \right)\right\}^{m^{\left(l\right)}_0}
		\cdots
		\left\{\theta^{\left({l}\right)}\left( s_{r-1} \right)\right\}^{m^{\left(l\right)}_{r-1}} 
	\right\}
	\right]
	\\[40pt]
	&=
		\displaystyle\sum_{\star 2}
		\left[
		\displaystyle\prod_{l=1}^{r-1}
		\left\{
		\dfrac{m^{\left(l\right) } !}{m^{\left(l\right)}_0 ! \cdots m^{\left(l\right)}_{r-1} !}
		\left( \left\{ \theta^{\left({l}\right)}\left( s_{0} \right)\right\} x_{0} \right)^{m^{\left(l\right)}_{0}}
		\cdots
		\left( \left\{ \theta^{\left({l}\right)}\left( s_{r-1} \right)\right\} x_{r-1} \right)^{m^{\left(l\right)}_{r-1}}
	\right\}
	\right]
	\\[30pt]
	&=
		\displaystyle\prod_{l=1}^{r-1}
		\left[
		\displaystyle\sum_{\star 2}
		\left\{
		\dfrac{m^{\left(l\right) } !}{m^{\left(l\right)}_0 ! \cdots m^{\left(l\right)}_{r-1} !}
		\left( \left\{ \theta^{\left({l}\right)}\left( s_{0} \right)\right\} x_{0} \right)^{m^{\left(l\right)}_{0}}
		\cdots
		\left( \left\{ \theta^{\left({l}\right)}\left( s_{r-1} \right)\right\} x_{r-1} \right)^{m^{\left(l\right)}_{r-1}}
	\right\}
	\right]
	\\[30pt]
	&=
		\displaystyle\prod_{l=1}^{r-1}
		\left(\displaystyle \sum_{k=0}^{r-1} \left\{\theta^{\left(l\right)} \left(k\right)\right\} \cdot x_k\right)^{m^{\left(l\right)}} 
	~=~
		\displaystyle\prod_{l=1}^{r-1}
		\left(y^{\left(l\right)}\right)^{m^{\left(l\right)}}.
	\end{split}
\end{eqnarray}
\end{widetext}

\subsection{ CDS: Exact formulation of lower-order moments for CDOS }
As shown, although we see that value of correlations for SDS conform to CDS at thermodynamic limit, whether and/or how their differences converge more rapidly (or slowly) than 
decrease of width of the configurational density of states (CDOS): If the former is not satisfied at certain conditions, SDS cannot reach majority of microscopic states corresponding to at (or near) CDS. 
Therefore, in order to confirm this convergence, we should first derive exact formulation of system-size dependence of 1-order and 2-order moment of CDOS (especially, for pair correlation) for multicomponent systems.
\subsubsection{Exact formulation of 1-order moments for products of any functions of spin variables}
Let us first consider $r$-component system with number of atoms $N$, where composition for component $X_k$ is respectively given by $x_k$.
Then we introduce any function (not necessarily basis function) of spin variable $\sigma_p$ at lattice point $p$, $\phi^{\left(l_p\right)} \left(\sigma_p\right)$. 
When we prepare such $q$ different functions $\left\{\phi^{\left(1\right)},\cdots,\phi^{\left(q\right)}\right\}$ where $\phi^{\left(l\right)}$ is defined on $m^{\left(l\right)}$ different lattice points, 1-order moment (i.e., trace over all symmetry-equivalent set of lattice point and over all possible configuration) of the product of these function on $m = \sum_{l=1}^{q} m^{\left(l\right)}$ lattice points can be given in analogy to the previous section by 
\begin{widetext}
\begin{eqnarray}
	\label{eq::exact_moment:XI_MOM1}
	\begin{split}
		&\hspace{-30pt}
			\Braket{
			\phi^{\left(1\right)} \left(\sigma_1\right)
			\cdots
			\phi^{\left(1\right)} \left(\sigma_{m^{\left(1\right)}}\right)
			\cdot
			\phi^{\left(2\right)} \left(\sigma_{m^{\left(1\right)}+1}\right)
			\cdots
			\phi^{\left(q\right)} \left(\sigma_{m}\right)
			}
		%[10pt]
		=
			\dfrac{\left(x_0 N\right) ! \cdots \left(x_{r-1} N\right) !}{N !}
			\displaystyle\sum_{\star 1}
			\left[
			\displaystyle\prod_{l=0}^{q}
			\left\{
			\dfrac{m^{\left(l\right)} !}{m^{\left(l\right)}_0 ! \cdots m^{\left(l\right)}_{r-1} !} 
			\left(
			\displaystyle\prod_{k=0}^{r-1}
			\left\{\phi^{\left(l\right)} \left(s_k\right) \right\}^{m^{\left(l\right)}_k}
			\right)
			\right\}
			\right],
	\end{split}
\end{eqnarray}
\end{widetext}
where $m^{\left(l\right)}_k$ denotes number of lattice points occupied by component $X_k$ with function of $\phi^{\left(l\right)}$, and $s_k$ represents value of spin variable for $X_k$. 
$\phi^{\left(0\right)}$ and $m^{\left(0\right)}$ are additionally introduced just for convenience, which satisfies $\phi^{\left(0\right)} = 1$ and $m^{\left(0\right)} = N - m$. 
Summation $\star 1$ is taken over $m^{\left(k\right)}_l$ satisfying the following:
\begin{eqnarray}
	\label{eq::exact_moment:star1}
		\star 1 
		~:~
			\displaystyle\sum_{l=0}^{q}
			m^{\left(l\right)}_k = x_k N,
			~\displaystyle\sum_{k=0}^{r-1} m_k^{\left(l\right)} = m^{\left(l\right)},
			~m^{\left(l\right)}_k \geq 0
\end{eqnarray}
For instance, when we explicitly choose a set of $\phi$ as a set of orthonormal basis function $\theta$, Eq.~(\ref{eq::exact_moment:XI_MOM1}) exactly corresponds to 1-order moment for 
conventional correlation functions. Eq.~(\ref{eq::exact_moment:XI_MOM1}) can be further simplified by dividing contribution from combination of $\phi^{\left(0\right)}$ from other functions: 
\begin{widetext}
\begin{eqnarray}
	\label{eq::exact_moment:XI_MOM2}
	\begin{split}
			&\displaystyle\sum_{\star 2}
			\left[
			\dfrac{\left(x_0 N\right) ! \cdots \left(x_{r-1} N\right) !}{N !}
			\dfrac{\left(N-m\right) !}{\left( x_0 N - \sum_{l=1}^{q} m^{\left(l\right)}_0 \right) ! \cdots \left( x_{r-1} N - \sum_{l=1}^{q} m^{\left(l\right)}_{r-1} \right) !}
		%\right.%\\[5pt]
		%&\hspace{150pt}\left.
			\cdot
			\displaystyle\prod_{l=1}^{q}
			\left\{
			\dfrac{m^{\left(l\right)} !}{m^{\left(l\right)}_0 ! \cdots m^{\left(l\right)}_{r-1} !}
			\left(
			\displaystyle\prod_{k=0}^{r-1}
			\left\{\phi^{\left(l\right)} \left(s_k\right) \right\}^{m^{\left(l\right)}_k}
			\right)
			\right\}
			\right],
	\end{split}
\end{eqnarray}
\end{widetext}
where summation $\star 2$ is taken for $m^{\left(l\right)}_k$ satisfying the following: 
\begin{eqnarray}
	\label{eq::exact_moment:star2}
	\star 2 
	~:~
		\displaystyle\sum_{l=1}^{q}
		m^{\left(l\right)}_k \leq m_k,
		~\displaystyle\sum_{k=0}^{r-1} m_k^{\left(l\right)} = m^{\left(l\right)},
		~m^{\left(l\right)}_k \geq 0.
\end{eqnarray}

Based on the above, we can now provide exact formulation of 1-order moment for product of a given set of $\phi$ on from 1 to 4 lattice points. 
For convenience, we first define the following linear combinations:
\begin{widetext}
\begin{eqnarray}
	\label{eq::exact_moment:PHI1}
	&\Phi^{\left(l\right)} =
		\sum_{k = 0}^{r-1} x_k \left\{\phi^{\left(l\right)} \left(s_k\right) \right\}
	&\\[10pt]
	\label{eq::exact_moment:PHI2}
	&\Phi^{\left(l_1,l_2\right)} =
		\sum_{k = 0}^{r-1} x_k \left\{\phi^{\left(l_1\right)} \left(s_k\right) \right\}
		\left\{\phi^{\left(l_2\right)} \left(s_k\right) \right\}
	&\\[10pt]
	\label{eq::exact_moment:PHI3}
	&\Phi^{\left(l_1,l_2,l_3\right)} = 
		\sum_{k = 0}^{r-1} x_k \left\{\phi^{\left(l_1\right)} \left(s_k\right) \right\}
		\left\{\phi^{\left(l_2\right)} \left(s_k\right) \right\}
		\left\{\phi^{\left(l_3\right)} \left(s_k\right) \right\}
	&\\[10pt]
	\label{eq::exact_moment:PHI4}
	&\Phi^{\left(l_1,l_2,l_3,l_4\right)} = 
		\sum_{k = 0}^{r-1} x_k \left\{\phi^{\left(l_1\right)} \left(s_k\right) \right\}
		\left\{\phi^{\left(l_2\right)} \left(s_k\right) \right\}
		\left\{\phi^{\left(l_3\right)} \left(s_k\right) \right\}
		\left\{\phi^{\left(l_4\right)} \left(s_k\right) \right\}&
\end{eqnarray}
\end{widetext}
Using these results, we can express 1-order moment for given $\phi^{\left(l\right)} \left(\sigma_1\right)$ at a given lattice point $\sigma_1$ as
\begin{widetext}
\begin{eqnarray}
	\label{eq::exact_moment:average_PHI1}
	\begin{split}
		\Braket{\phi^{\left(l\right)} \left(\sigma_1\right)}
		&= 	
		\begin{split}
				&\displaystyle\sum_{\star 2}
				\left[
				\dfrac{\left(x_0 N\right) ! \cdots \left(x_{r-1} N\right) !}{N !}
				\dfrac{\left(N-1\right) !}{\left( x_0 N - m^{\left(l\right)}_0 \right) ! \cdots \left( x_{r-1} N - m^{\left(l\right)}_{r-1} \right) !}
				\right.%\\[5pt]
			%\hspace{100pt}
			\left.
				\cdot
				\left\{
				\dfrac{1!}{m^{\left(l\right)}_0 ! \cdots m^{\left(l\right)}_{r-1} !}
				\left(
				\displaystyle\prod_{k=0}^{r-1} \left\{\phi^{\left(l\right)} \left(s_k\right) \right\}^{m^{\left(l\right)}_k}
				\right)
				\right\}
				\right]
		\end{split}\\	
		&=\displaystyle\sum_{k = 0}^{r-1}
			\left[ x_k \left\{\phi^{\left(l\right)} \left(s_k\right) \right\}\right]
			=
			\Phi^{\left(l\right)}
	\end{split}
\end{eqnarray}
\end{widetext}
In a similar fashion, 1-order moment for products of functions on from 2 to 4 lattice points can be given by
\newpage
\begin{widetext}
\begin{eqnarray}
	\label{eq::exact_moment:average_PHI2}
	&\Braket{
	\left\{\phi^{\left(l_1\right)} \left(\sigma_1\right)\right\}
	\left\{\phi^{\left(l_2\right)} \left(\sigma_2\right)\right\}
	}=
		\dfrac{N \Phi^{\left(l_1\right)} \Phi^{\left(l_2\right)} - \Phi^{\left(l_1,l_2\right)}}
		{N - 1}
	&\\[20pt]
	\label{eq::exact_moment:average_PHI3}
	&\Braket{
	\left\{\phi^{\left(l_1\right)} \left(\sigma_1\right)\right\}
	\left\{\phi^{\left(l_2\right)} \left(\sigma_2\right)\right\}
	\left\{\phi^{\left(l_3\right)} \left(\sigma_2\right)\right\}
	}=
		\dfrac{
		N^2 \Phi^{\left(l_1\right)} \Phi^{\left(l_2\right)} \Phi^{\left(l_3\right)}
		-
		N \sum_j \Phi^{\left(j_1\right)} \Phi^{\left(j_2,j_3\right)}
		+2 \Phi^{\left(l_1,l_2,l_3\right)}}{\left(N - 1\right) \left(N - 2\right)}
	&\\[20pt]
	\label{eq::exact_moment:average_PHI4}
	&\begin{split}
		&\Braket{
		\left\{\phi^{\left(l_1\right)} \left(\sigma_1\right)\right\}
		\left\{\phi^{\left(l_2\right)} \left(\sigma_2\right)\right\}
		\left\{\phi^{\left(l_3\right)} \left(\sigma_2\right)\right\}
		\left\{\phi^{\left(l_4\right)} \left(\sigma_2\right)\right\}}
		=
			\dfrac{1}{\left(N - 1\right) \left(N - 2\right) \left(N - 3\right)}
			\left[
			N^3 \Phi^{\left(l_1\right)} \Phi^{\left(l_2\right)} \Phi^{\left(l_3\right)} \Phi^{\left(l_4\right)}
		\right.\\
		&\left.
			-
			N^2 \sum_j \Phi^{\left(j_1\right)} \Phi^{\left(j_2\right)} \Phi^{\left(j_3,j_4\right)}
			+
			N \left\{2 \sum_j \Phi^{\left(j_1\right)} \Phi^{\left(j_2,j_3,j_4\right)}
			+
			\sum_j \Phi^{\left(j_1,j_2\right)} \Phi^{\left(j_3,j_4\right)}  \right\}
			-
			6\Phi^{\left(j_1,j_2,j_3,j_4\right)}
			\right], 
	\end{split}&
\end{eqnarray}
\end{widetext}
where summation $j$ is taken for all possible combination of a set of $l$ excluding their duplicate permutation, e.g., $\Phi^{\left(l_1,l_2\right)}$ and $\Phi^{\left(l_2,l_1\right)}$. 

\subsubsection{Formulation of 1- and 2-order moments on binary system}
Now we are ready to paractically provide exact formulation of 1- and 2-order moments for point and pair correlations on binary and ternary system with conventional basis functions, where 
1- and 2-order moments on binary system specific to certain basis functions has been given by our recent study.
We first prepare spin variable $\sigma_p=+1$ for A and $\sigma_p=-1$ for B occupation for binary A$_{\left(1-x\right)}$B$_x$ system. 
Then conventional, orthonormal basis functions for generalized Ising system can be 
obtained by applying Gram-Schmidt technique to linearly independent functions of spin variable, which leads to 
\begin{eqnarray}
	\label{eq::exact_moment:THETA_binary}
	\theta^{\left(0\right)} \left(\sigma_p\right) = 1,
	~\theta^{\left(1\right)} \left(\sigma_p\right) = \sigma_p.
\end{eqnarray}
Then we define function $\phi$ as 
\begin{eqnarray}
	\label{eq::exact_moment:PHI_binary}
	\phi^{\left(1\right)} \left(\sigma_p\right) = \theta^{\left(1\right)} \left(\sigma_p\right),
	~\phi^{\left(11\right)} \left(\sigma_p\right) = \left\{\theta^{\left(1\right)}\right\}^2 \left(\sigma_p\right).
\end{eqnarray}
Hereinafter $m_s$ denotes a figure composed of lattice points in the system, where $m$ and $s$ respectively specifies the number of lattice points included in the figure and the shape of figure.
When we also define $y=1-2x$, 1-order moment of point correlation can be immediately given by
\begin{eqnarray}
	\label{eq::exact_moment:average_1body_binary}
	\Braket{\xi_{1}} = \Braket{\phi^{\left(1\right)} \left(\sigma_1\right)} = 1-2x := y, 
\end{eqnarray}
and 1-order moment for pair correlation can be given by
\begin{eqnarray}
	\label{eq::exact_moment:average_2body_binary}
	\Braket{\xi_{2}} = \Braket{\left\{\phi^{\left(1\right)} \left(\sigma_1\right)\right\}\left\{\phi^{\left(1\right)} \left(\sigma_2\right)\right\}} = \dfrac{N y^2 - 1}{N-1}.
\end{eqnarray}
Particularly at equiatomic composition, this can be simplified to
\begin{eqnarray}
	\label{eq::exact_moment:average_2body_binary_eq}
	\Braket{\xi_{2}} = -\dfrac{1}{N-1}.
\end{eqnarray}
In a similar fashion, 2-order moment for pair correlation is given by
\begin{widetext}
\begin{eqnarray}
	\label{eq::exact_moment:2ndmom_2body_binary}
	\begin{split}
		&\Braket{\xi_{2_s}^2}
		=
			\dfrac{1}{\left(D_s N\right)^2}
			\left\{
			F
			\begin{pmatrix}
				\circ\hspace{-1pt}{\textendash }\hspace{-1pt}\circ\\\vspace{-15pt}~\vspace{-15pt}\\\circ\hspace{-1pt}{\textendash}\hspace{-1pt}\circ
			\end{pmatrix}
			\Braket{
			\left\{\phi^{\left(11\right)} \left(\sigma_1\right)\right\}
			\left\{\phi^{\left(11\right)} \left(\sigma_2\right)\right\}
			}
		\right.\\
		&\hspace{80pt}\left.
			+F
			\begin{pmatrix}
				\circ\hspace{-1pt}{\textendash }\hspace{-1pt}\circ\hspace{8pt}\\\vspace{-15pt}~\vspace{-15pt}\\\hspace{8pt}\circ\hspace{-1pt}{\textendash}\hspace{-1pt}\circ
			\end{pmatrix}
			\Braket{
			\left\{\phi^{\left(1\right)} \left(\sigma_1\right)\right\} \left\{\phi^{\left(11\right)}
			\left(\sigma_2\right)\right\}\left\{\phi^{\left(1\right)} \left(\sigma_3\right)\right\}}
		\right.\\
		&\hspace{120pt}\left.
			+F
			\begin{pmatrix}
				\circ\hspace{-1pt}{\textendash }\hspace{-1pt}\circ\hspace{4pt}\\\vspace{-15pt}~\vspace{-15pt}\\\hspace{4pt}\circ\hspace{-1pt}{\textendash}\hspace{-1pt}\circ
			\end{pmatrix}
			\Braket{
			\left\{\phi^{\left(1\right)} \left(\sigma_1\right)\right\} \left\{\phi^{\left(1\right)} \left(\sigma_2\right)\right\}\left\{\phi^{\left(1\right)} \left(\sigma_3\right)\right\} \left\{\phi^{\left(1\right)} \left(\sigma_4\right)\right\}
			}
			\right\}
		\\[5pt]
		&\hspace{10pt}=
			\dfrac {1}{D_{s} \left( N-1 \right)  \left( N-2 \right)  \left( N-3 \right) }
			\left\{{y}^{4}D_{s} {N}^{3}
			+
			\left( -4 D_{s} {y}^{4}+{y}^{4}-2 D_{s} {y}^{2}-2 {y}^{2}+1 \right){N}^{2}
		\right.\\
		&\hspace{50pt}\left.
			+
			\left( 12 D_{s} {y}^{2}+4 {y}^{2}-D_{s}-4 \right) N
			-
			8 D_{s} {y}^{2}-4 {y}^{2}+2 D_{s}+4
			\right\},
	\end{split}
\end{eqnarray}
\end{widetext}
where $F$ denotes number of diagrams consisting of two symmetry-equivalent pairs to considered pair, whose argument denotes 
number of shared lattice points between the two pairs. 
\begin{comment}
考えてる図形がセル内に収まる十分大きい$N$ではその式の形は変わらない．
今回は十分大きいセルに対して計算している．
（例えば最近接二体の場合, 
プリミティブセルを3倍以上ずつしたセルでは$F$の式は変わらないが, 
それより小さいセルに対しては$F$の形が変わるため注意が必要である）．
\end{comment}
For equiatomic composition, this can also be simplified to
\begin{eqnarray}
	\label{eq::exact_moment:2ndmom_2body_binary_eq}
	\Braket{\xi_{2_s}^2} = {\dfrac {N- \left(D_{s}+2\right)}{ D_{s} \left( N-3 \right) \left( N-1 \right) }}.
\end{eqnarray}

\subsubsection{Formulation of 1- and 2-order moments on ternary system}
In a similar fashion to binary system, we prepare spin variable $\sigma_p=+1,0,-1$ for A, B and C occupation for ternary system. 
Then conventional, orthonormal basis functions can be given by 
\begin{eqnarray}
	\label{eq::exact_moment:THETA_ternary}
	\theta^{\left(0\right)} \left(\sigma_p\right) = 1,
	~\theta^{\left(1\right)} \left(\sigma_p\right) = \dfrac{\sqrt{6}}{2}\sigma_p,
	~\theta^{\left(2\right)} \left(\sigma_p\right) = \dfrac{3\sqrt{2}}{2}\sigma_p^2 - \sqrt{2}. \nonumber \\
	\quad
\end{eqnarray}
Then functions $\phi$ are defined as
\begin{widetext}
\begin{eqnarray}
	\label{eq::exact_moment:PHI_ternary}
	\begin{split}
		&\hspace{100pt}
			\phi^{\left(1\right)} \left(\sigma_p\right) = \theta^{\left(1\right)} \left(\sigma_p\right),
			~\phi^{\left(2\right)} \left(\sigma_p\right) = \theta^{\left(2\right)} \left(\sigma_p\right)
		\\[5pt]
			&\phi^{\left(11\right)} \left(\sigma_p\right) = \left\{\theta^{\left(1\right)}\right\}^2 \left(\sigma_p\right),
			~\phi^{\left(22\right)} \left(\sigma_p\right) = \left\{\theta^{\left(2\right)}\right\}^2 \left(\sigma_p\right),
			~\phi^{\left(12\right)} \left(\sigma_p\right) = \left\{\theta^{\left(1\right)}\right\}\left\{\theta^{\left(2\right)}\right\} \left(\sigma_p\right).
	\end{split}
\end{eqnarray}
\end{widetext}
Hereinafter $m_s \left(i_1,\cdots,i_m\right)$ denotes a figure composed of lattice points in the system, where $m$ and $s$ respectively specifies the number of lattice points included in the figure and the shape of figure, and specific to ternary (as well as multicomponent) system, $\left(i_1,\cdots,i_m\right)$ denotes a set of basis function index included in the considered figure. 
Then 1-order moment for point correlatioins are given by 
\begin{eqnarray}
	\label{eq::exact_moment:average_1body_ternary1}
	&\Braket{\xi_{1 \left(1\right)}} = \Braket{\phi^{\left(1\right)} \left(\sigma_1\right)} = \dfrac{\sqrt {6} \left( 1-x_{\rm B}-2 x_{\rm C} \right)}{2} := y_1&\\[10pt]
	\label{eq::exact_moment:average_1body_ternary2}
	&\Braket{\xi_{1 \left(2\right)}} = \Braket{\phi^{\left(2\right)} \left(\sigma_1\right)} = \dfrac{\sqrt {2} \left( 1-3 x_{\rm B} \right)}{2} := y_2&, 
\end{eqnarray}
and 1-order moment for pair correlations are given by
\begin{widetext}
\begin{eqnarray}
	\label{eq::exact_moment:average_2body_ternary11}
	&\Braket{\xi_{2 \left(1,1\right)}}= 
		\Braket{
		\left\{\phi^{\left(1\right)} \left(\sigma_1\right)\right\}
		\left\{\phi^{\left(1\right)} \left(\sigma_2\right)\right\}
		}
	=
		\dfrac {2 N{y_{1}}^{2}-\sqrt {2}y_{2}-2}{2 N-2}&\\[10pt]
	\label{eq::exact_moment:average_2body_ternary22}
	&\Braket{\xi_{2 \left(2,2\right)}}=
		\Braket{
		\left\{\phi^{\left(2\right)} \left(\sigma_1\right)\right\}
		\left\{\phi^{\left(2\right)} \left(\sigma_2\right)\right\}
		}
	=
		\dfrac {2 N{y_{2}}^{2}+\sqrt {2}y_{2}-2}{2 N-2}&\\[10pt]
	\label{eq::exact_moment:average_2body_ternary12}
	&\Braket{\xi_{2 \left(1,2\right)}}=
		\Braket{
		\left\{\phi^{\left(1\right)} \left(\sigma_1\right)\right\}
		\left\{\phi^{\left(2\right)} \left(\sigma_2\right)\right\}
		}
	=
		\dfrac { \left( 2 Ny_{2}-\sqrt {2} \right) y_{1}}{2 N-2}&, 
\end{eqnarray}
\end{widetext}
where at equiatomic composition, this can be rewritten as 
\begin{eqnarray}
	\label{eq::exact_moment:average_2body_ternary_eq}
	\Braket{\xi_{2 \left(1,1\right)}} = \Braket{\xi_{2 \left(2,2\right)}} = -\dfrac{1}{N-1},
	~~~\Braket{\xi_{2 \left(1,2\right)}} = 0. 
\end{eqnarray}
When extending the formulation for binary system above, 2-order moment for pair correlations are given by 
\begin{widetext}
\begin{eqnarray}
	\label{eq::exact_moment:2ndmom_2body_ternary11}
	&\begin{split}
	&\Braket{\xi_{2_s \left(1,1\right)}^2}=
		\dfrac{1}{\left(D_s N\right)^2}
		\left\{
		F
		\begin{pmatrix}
			\circ\hspace{-1pt}{\textendash }\hspace{-1pt}\circ\\\vspace{-15pt}~\vspace{-15pt}\\\circ\hspace{-1pt}{\textendash}\hspace{-1pt}\circ
		\end{pmatrix}
		\Braket{
		\left\{\phi^{\left(11\right)} \left(\sigma_1\right)\right\}
		\left\{\phi^{\left(11\right)} \left(\sigma_2\right)\right\}
		}
	\right.\\
	&\hspace{80pt}\left.
		+F
		\begin{pmatrix}
			\circ\hspace{-1pt}{\textendash }\hspace{-1pt}\circ\hspace{8pt}\\\vspace{-15pt}~\vspace{-15pt}\\\hspace{8pt}\circ\hspace{-1pt}{\textendash}\hspace{-1pt}\circ
		\end{pmatrix}
		\Braket{
		\left\{\phi^{\left(1\right)} \left(\sigma_1\right)\right\}
		\left\{\phi^{\left(11\right)} \left(\sigma_2\right)\right\}
		\left\{\phi^{\left(1\right)} \left(\sigma_3\right)\right\}
		}
	\right.\\
	&\hspace{120pt}\left.
		+F
		\begin{pmatrix}
			\circ\hspace{-1pt}{\textendash }\hspace{-1pt}\circ\hspace{4pt}\\\vspace{-15pt}~\vspace{-15pt}\\\hspace{4pt}\circ\hspace{-1pt}{\textendash}\hspace{-1pt}\circ
		\end{pmatrix}
		\Braket{
		\left\{\phi^{\left(1\right)} \left(\sigma_1\right)\right\}
		\left\{\phi^{\left(1\right)} \left(\sigma_2\right)\right\}
		\left\{\phi^{\left(1\right)} \left(\sigma_3\right)\right\}
		\left\{\phi^{\left(1\right)} \left(\sigma_4\right)\right\}
		}
		\right\}
	\\[5pt]
	&\hspace{10pt}
		=
		\dfrac{1}{4 D_{s} \left( N-1 \right)  \left( N-2 \right)  \left( N-3 \right) }
		\left\{ 4 {y_{1}}^{4}D_{s} {N}^{3}
	\right.\\
	&\hspace{10pt}\left.
		+ \left( -4 D_{s} \sqrt {2}{y_{1}}^{2}y_{2}-16 D_{s} {y_{1}}^{4}-4 {y_{1}}^{2}\sqrt {2}y_{2}+4 {y_{1}}^{4}-8 D_{s} {y_{1}}^{2}+4 \sqrt {2}y_{2}-8 {y_{1}}^{2}+2 {y_{2}}^{2}+4 \right) {N}^{2} 
	\right.\\
	&\hspace{10pt}\left.
		+\left( 24 D_{s} \sqrt {2}{y_{1}}^{2}y_{2}-4 D_{s} \sqrt {2}y_{2}+48 D_{s} {y_{1}}^{2}-2 D_{s} {y_{2}}^{2}-15 \sqrt {2}y_{2}+24 {y_{1}}^{2}-6 {y_{2}}^{2}-4 D_{s}-18 \right) N
	\right.\\
	&\left.\hspace{10pt}
		+6 D_{s} \sqrt {2}y_{2}-48 D_{s} {y_{1}}^{2}+15 \sqrt {2}y_{2}-24 {y_{1}}^{2}+6 {y_{2}}^{2}+12 D_{s}+18
	\right\}
	\end{split}&\\[20pt]
	\label{eq::exact_moment:2ndmom_2body_ternary22}
	&\begin{split}
		&\Braket{\xi_{2_s \left(2,2\right)}^2}=
			\dfrac{1}{\left(D_s N\right)^2}
			\left\{
			F
			\begin{pmatrix}
				\circ\hspace{-1pt}{\textendash }\hspace{-1pt}\circ\\\vspace{-15pt}~\vspace{-15pt}\\\circ\hspace{-1pt}{\textendash}\hspace{-1pt}\circ
			\end{pmatrix}
			\Braket{
			\left\{\phi^{\left(22\right)} \left(\sigma_1\right)\right\}
			\left\{\phi^{\left(22\right)} \left(\sigma_2\right)\right\}
			}
		\right.\\
	&\hspace{80pt}\left.
		+F
		\begin{pmatrix}
			\circ\hspace{-1pt}{\textendash }\hspace{-1pt}\circ\hspace{8pt}\\\vspace{-15pt}~\vspace{-15pt}\\\hspace{8pt}\circ\hspace{-1pt}{\textendash}\hspace{-1pt}\circ
		\end{pmatrix}
		\Braket{
		\left\{\phi^{\left(2\right)} \left(\sigma_1\right)\right\}
		\left\{\phi^{\left(22\right)} \left(\sigma_2\right)\right\}
		\left\{\phi^{\left(2\right)} \left(\sigma_3\right)\right\}
		}
	\right.\\
	&\hspace{120pt}\left.
		+F
		\begin{pmatrix}
			\circ\hspace{-1pt}{\textendash }\hspace{-1pt}\circ\hspace{4pt}\\\vspace{-15pt}~\vspace{-15pt}\\\hspace{4pt}\circ\hspace{-1pt}{\textendash}\hspace{-1pt}\circ
		\end{pmatrix}
		\Braket{
		\left\{\phi^{\left(2\right)} \left(\sigma_1\right)\right\}
		\left\{\phi^{\left(2\right)} \left(\sigma_2\right)\right\}
		\left\{\phi^{\left(2\right)} \left(\sigma_3\right)\right\}
		\left\{\phi^{\left(2\right)} \left(\sigma_4\right)\right\}
		}
		\right\}
	\\[5pt]
	&\hspace{10pt}
		=
			\dfrac {1}{4 D_{s}  \left( N-1 \right)  \left( N-2 \right)  \left( N-3 \right) }
			\left\{ 4 {y_{2}}^{4}D_{s} {N}^{3} 
		\right.\\
	&\hspace{10pt}\left.
		+\left( 4 D_{s} \sqrt {2}{y_{2}}^{3}-16 D_{s} {y_{2}}^{4}+4 {y_{2}}^{3}\sqrt {2}+4 {y_{2}}^{4}-8 D_{s} {y_{2}}^{2}-4 \sqrt {2}y_{2}-6 {y_{2}}^{2}+4 \right) {N}^{2}
	\right.\\
	&\hspace{10pt}\left.
		+\left( -24 D_{s} \sqrt {2}{y_{2}}^{3}+4 D_{s} \sqrt {2}y_{2}+46 D_{s} {y_{2}}^{2}+9 \sqrt {2}y_{2}+18 {y_{2}}^{2}-4 D_{s}-18 \right) N
	\right.\\
	&\hspace{10pt}\left.
		+6 D_{s} \sqrt {2}y_{2}-48 D_{s} {y_{2}}^{2}-9 \sqrt {2}y_{2}-18 {y_{2}}^{2}+12 D_{s}+18
		\right\}
	\end{split}&\\[20pt]
	\label{eq::exact_moment:2ndmom_2body_ternary12}
	&\begin{split}
	&\Braket{\xi_{2_s \left(1,2\right)}^2}=
			\dfrac{1}{\left(2 D_s N\right)^2}
			\left\{
			F
			\begin{pmatrix}
				\circ\hspace{-1pt}{\textendash }\hspace{-1pt}\circ\\\vspace{-15pt}~\vspace{-15pt}\\\circ\hspace{-1pt}{\textendash}\hspace{-1pt}\circ
			\end{pmatrix}
			\left(
			2\Braket{
			\left\{\phi^{\left(11\right)} \left(\sigma_1\right)\right\}
			\left\{\phi^{\left(22\right)} \left(\sigma_2\right)\right\}
			}
			+2\Braket{
			\left\{\phi^{\left(12\right)} \left(\sigma_1\right)\right\}
			\left\{\phi^{\left(12\right)} \left(\sigma_2\right)\right\}
			}
			\right)
		\right.\\
		&\hspace{10pt}\left.
			+F
			\begin{pmatrix}
				\circ\hspace{-1pt}{\textendash }\hspace{-1pt}\circ\hspace{8pt}\\\vspace{-15pt}~\vspace{-15pt}\\\hspace{8pt}\circ\hspace{-1pt}{\textendash}\hspace{-1pt}\circ
			\end{pmatrix}
			\left(
			2\Braket{
			\left\{\phi^{\left(1\right)} \left(\sigma_1\right)\right\}
			\left\{\phi^{\left(12\right)} \left(\sigma_2\right)\right\}
			\left\{\phi^{\left(2\right)} \left(\sigma_3\right)\right\}}
			\right.\right.\\
		&\hspace{30pt}\left.\left.
			+\Braket{
			\left\{\phi^{\left(1\right)} \left(\sigma_1\right)\right\}
			\left\{\phi^{\left(22\right)} \left(\sigma_2\right)\right\}
			\left\{\phi^{\left(1\right)} \left(\sigma_3\right)\right\}
			}
			+
			\Braket{
			\left\{\phi^{\left(2\right)} \left(\sigma_1\right)\right\}
			\left\{\phi^{\left(11\right)} \left(\sigma_2\right)\right\}
			\left\{\phi^{\left(2\right)} \left(\sigma_3\right)\right\}
			}
			\right)
		\right.\\
		&\hspace{120pt}\left.
			+4F
			\begin{pmatrix}
				\circ\hspace{-1pt}{\textendash }\hspace{-1pt}\circ\hspace{4pt}\\\vspace{-15pt}~\vspace{-15pt}\\\hspace{4pt}\circ\hspace{-1pt}{\textendash}\hspace{-1pt}\circ
			\end{pmatrix}
			\Braket{
			\left\{\phi^{\left(1\right)} \left(\sigma_1\right)\right\}
			\left\{\phi^{\left(2\right)} \left(\sigma_2\right)\right\}
			\left\{\phi^{\left(1\right)} \left(\sigma_3\right)\right\}
			\left\{\phi^{\left(2\right)} \left(\sigma_4\right)\right\}
			}
			\right\}
		\\[5pt]
		&\hspace{10pt}
			=\dfrac {1}{4 D_{s}  \left( N-1 \right)  \left( N-2 \right)  \left( N-3 \right) }
			\left\{ 4 {y_{1}}^{2}{y_{2}}^{2}D_{s} {N}^{3}
		\right.\\
		&\hspace{10pt}\left.
			+
			\left( -4 D_{s} \sqrt {2}{y_{1}}^{2}y_{2}-16 D_{s} {y_{1}}^{2}{y_{2}}^{2}-{y_{1}}^{2}\sqrt {2}y_{2}-{y_{2}}^{3}\sqrt {2}+4 {y_{1}}^{2}{y_{2}}^{2}-{y_{1}}^{2}-3 {y_{2}}^{2}+2 \right) {N}^{2}
		\right.\\
		&\hspace{10pt}\left.
			+
			\left( 18 D_{s} \sqrt {2}{y_{1}}^{2}y_{2}+2 D_{s} \sqrt {2}{y_{2}}^{3}-3 {y_{1}}^{2}\sqrt {2}y_{2}+{y_{2}}^{3}\sqrt {2}
		\right.\right.\\
		&\left.\left.\hspace{100pt}
			+
			4 D_{s} {y_{1}}^{2}+6 D_{s} {y_{2}}^{2}+3 \sqrt {2}y_{2}+3 {y_{1}}^{2}+9 {y_{2}}^{2}-4 D_{s}-8 \right) N
		\right.\\
		&\hspace{10pt}\left.
			-6 D_{s} \sqrt {2}y_{2}-12 D_{s} {y_{1}}^{2}-12 D_{s} {y_{2}}^{2}-3 \sqrt {2}y_{2}-6 {y_{2}}^{2}+12 D_{s}+6
			\right\},
		\end{split}&
\end{eqnarray}
\end{widetext}
where at equiatomic composition, this can be simplified to 
\begin{eqnarray}
	\label{eq::exact_moment:2ndmom_2body_ternary11_eq}
	&\Braket{\xi_{2_s \left(1,1\right)}^2} = \dfrac{2 N-\left(2 D_{s}+3\right)}{2 D_{s}  \left( N-1 \right)  \left( N-2\right)}&\\[10pt]
	\label{eq::exact_moment:2ndmom_2body_ternary22_eq}
	&\Braket{\xi_{2_s \left(2,2\right)}^2} = \dfrac{2 N-\left(2 D_{s}+3\right)}{2 D_{s}  \left( N-1 \right)  \left( N-2\right)}&\\[10pt]
	\label{eq::exact_moment:2ndmom_2body_ternary12_eq}
	&\Braket{\xi_{2_s \left(1,2\right)}^2} = \dfrac{ N-\left(2 D_{s}+1\right)}{2 D_{s}  \left( N-1 \right)  \left( N-2\right)}&.
\end{eqnarray}

\subsection{SDS and CDS: Difference w.r.t. Landscape of CDOS}
Using all the results discussed above, we can finally estimate the difference in pair correlations between SDS and CDS w.r.t. the width of CDOS, as functions of system size and geometric character of underlying lattice, $\epsilon^2$, defined as 
\begin{eqnarray}
	\varepsilon^2 = 
		\left(\dfrac{\xi_{\rm{SDS}} - \Braket{\xi_{\textrm{CDS}}}}
		{\sqrt{\Braket{\xi_{\textrm{CDS}}^2} - \Braket{\xi_{\textrm{CDS}}}^2}}\right)^2.
\end{eqnarray}
Detailed derivation of the landscape of $\epsilon^2$ and its maximum and minimum values is significantly complicated, which therefore is 
given in Appendix. 

For binary system, minimum and maximum value of $\epsilon_1^2$ at thermodynamic limit are given by
\begin{eqnarray}
		&{\rm min}\left( \varepsilon_1^2 \right) \sim \dfrac{D_s}{N}
		,~
		\left( x_{\rm A},x_{\rm B} \right) = \left( \dfrac{1}{2},~\dfrac{1}{2} \right)
		&\\[5pt]
		&{\rm max}\left( \varepsilon_1^2 \right) \sim \dfrac{2 D_s}{N}
		,~
		\left( x_{\rm A},x_{\rm B} \right) \sim \left( 1,0 \right),\left( 0,~1 \right)
		&
\end{eqnarray}

For ternary system, we have three kinds of $\epsilon^2$, corresponding to basis index set of $\left(11\right)$, $\left(22\right)$ and $\left(12\right)$. 
At thermodynamic limit and for index set of $\left(11\right)$, we have 
\begin{widetext}
\begin{eqnarray}
	&{\rm min}\left( \varepsilon_{11}^2 \right) \sim \dfrac{D_s}{N},~\left(x_{\rm A},x_{\rm B},x_{\rm C}\right) \sim 
	\left(0,~\dfrac{1}{2},~\dfrac{1}{2} \right),
	\left(~\dfrac{1}{2},~0,~\dfrac{1}{2} \right),
	\left(\dfrac{1}{2},~\dfrac{1}{2},~0 \right)&\\[5pt]
	&{\rm max}\left( \varepsilon_{11}^2 \right) \sim \dfrac{50 D_s}{33N},~\left(x_{\rm A},x_{\rm B},x_{\rm C}\right) \sim 
	\left(1,~0,~0 \right),
	\left(0,~0,~1 \right),
\end{eqnarray}
\end{widetext}
for set of $\left(2\right)$, we have
\begin{eqnarray}
	&{\rm min}\left( \varepsilon_{22}^2 \right) \sim \dfrac{D_s}{N},~x_{\rm B} = \dfrac{1}{2}&\\[5pt]
	&{\rm max}\left( \varepsilon_{22}^2 \right) \sim \dfrac{2 D_s}{N},~x_{\rm B} \sim 0&,
\end{eqnarray}
and for set of $\left(12\right)$, it is given by
\begin{eqnarray}
	&{\rm min}\left( \varepsilon_{12}^2 \right) = 0,~x_{\rm A} = x_{\rm C}&\\[5pt]
	&{\rm max}\left( \varepsilon_{12}^2 \right) \sim \dfrac{D_s}{N},~\left(x_{\rm A},x_{\rm B},x_{\rm C}\right) \sim 
	\left( \dfrac{1}{2},~\dfrac{1}{2},~0 \right),
	\left( 0,~\dfrac{1}{2},~\dfrac{1}{2} \right). \nonumber \\
	\quad
\end{eqnarray}
From above results, we can clearly see that differences in pair correlation, $\epsilon^2$, between SDS and CDS is maximally the order of 
$\sim D_s/N$ for all pairs on binary as well as ternary system, which exhibits convergence at thermodynamic limit. 

We finally note that practically important points are (i) for a moderate system size typically found in e.g., treating alloy nanoparticles and locally-equilibrium region in subsystem including surface and interface, the difference in SDS and CDS for pair correlations still remains several percents for \textit{whole} composition, and (ii) significant $D$- and $x$-dependence of the deviations severely enhance error in constructing CDS based on the concept of SDS especially for the system where coordination number depends on the spatial position, e.g., nanoclusters and bumpy surfaces with defects. For such systems, CDS should be constructed based on our derived expressions.

\section{Conclusions}
Difference in stochastically disordered state (SDS) and configurationally disordered state (CDS) for discrete system under constant composition is quantitatively investigated based on exact expression of 1- and 2-order moments for configurational density of states (CDOS). We show that although SDS conforms to CDS at thermodynamic limit for any composition, there remains non-negligible difference between SDS and CDS (above few percents) for a moderate system size of a several ten thouthands of atoms, whose deviation significantly depends on the geometry of the figure in the system.

\section{Acknowledgement}
This work was supported by Grant-in-Aids for Scientific Research on Innovative Areas on High Entropy Alloys through the grant number JP18H05453 and a Grant-in-Aid for Scientific Research (16K06704) from the MEXT of Japan, Research Grant from Hitachi Metals$\cdot$Materials Science Foundation, and Advanced Low Carbon Technology Research and Development Program of the Japan Science and Technology Agency (JST).

\section*{Appendix: Derivation of maximum and minimum value for $\epsilon^2$}
We here show detailed derivation of differences between SDS and CDS w.r.t. width of CDOS, $\epsilon^2$.
\subsection*{Binary system}
For binary system, we consider large system where our derivation shown above exactly holds, which just satisfies $N\ge 27$ and number of A and B atoms are both greater or equal to 2. 
With these preparations, we introduce variable $y$ as 
\begin{eqnarray}
	\label{eq::epsilon:yrange_2c}
		&y = 1-2x_{\rm B}\\[5pt]
		&-1+\dfrac{4}{N} \leq y \leq 1-\dfrac{4}{N}.
\end{eqnarray}
Using $y$ and formulation for 1- and 2-order moments, $\varepsilon_1^2$ for considered pair $2_s$ is given by
\begin{widetext}
\begin{eqnarray}
	\label{eq::epsilon:1_ep^2}
	\varepsilon_1^2
	=
	\dfrac 
		{D_{s} \left( N-2 \right) \left( N-3 \right) }
		{N-2 D_{s}-1}
	\cdot
	\underbrace{
		\dfrac
			{\left( 1-y \right)  \left( 1+y \right)}
			{\left( N\left( 1-y \right) -2 \right)  \left( N \left( 1+y \right) -2\right)}
	}_{f}
\end{eqnarray}
\end{widetext}
Differenciation of $f$ in Eq.~(\ref{eq::epsilon:1_ep^2}) w.r.t. $y$ leads to
\begin{eqnarray}
	\label{eq::epsilon:1_f^(1)}
	\dfrac{\partial f}{\partial y}
	=
	\dfrac
		{8 \left( N-1 \right) y}
		{\left( Ny-N+2 \right) ^{2} \left( Ny+N-2 \right) ^{2}}. 
\end{eqnarray}
Therefore, $\varepsilon_1^2 \left(y\right)$ takes local minimum at $y=0$, and can take maximum at $y=-1+4/N,~1-4/N$.
Their individual values are given by 
\begin{eqnarray}
	\label{eq::epsilon2:1_ep_ex1}
	&\dfrac
			{D_{s} \left( N-3 \right)}
			{\left( N-2D_{s}-1 \right) \left( N-2 \right) }
	,~
	\left( x_{\rm A},x_{\rm B} \right) = \left( \dfrac{1}{2},\dfrac{1}{2} \right)
	&
	\\[5pt]
	\label{eq::epsilon2:1_ep_ex2}
	&\dfrac
			{2 D_{s} \left( N-2 \right)^{2}}
			{{N}^{2} \left( N-2D_{s}-1\right)}
	,~
	\left( x_{\rm A},x_{\rm B} \right) = \left( \dfrac{2}{N},~1-\dfrac{2}{N} \right)
		,~\left( 1-\dfrac{2}{N},~\dfrac{2}{N} \right) \nonumber \\
		\quad
	%&
\end{eqnarray}
From Eqs.~\eqref{eq::epsilon2:1_ep_ex1} and~\eqref{eq::epsilon2:1_ep_ex2}, 
minimum and maximum value of $\varepsilon_1^2$ at thermodynamic limit becomes
\begin{eqnarray}
		&{\rm min}\left( \varepsilon_1^2 \right) \sim \dfrac{D_s}{N}
		,~
		\left( x_{\rm A},x_{\rm B} \right) = \left( \dfrac{1}{2},~\dfrac{1}{2} \right)
		&\\[5pt]
		&{\rm max}\left( \varepsilon_1^2 \right) \sim \dfrac{2 D_s}{N}
		,~
		\left( x_{\rm A},x_{\rm B} \right) \sim \left( 1,0 \right),\left( 0,~1 \right).
		&
\end{eqnarray}

\subsection*{Ternary system}
\subsubsection{Basis index set of $\left(11\right)$}

\begin{figure}
	\centering
	\includegraphics[width=0.75\linewidth]{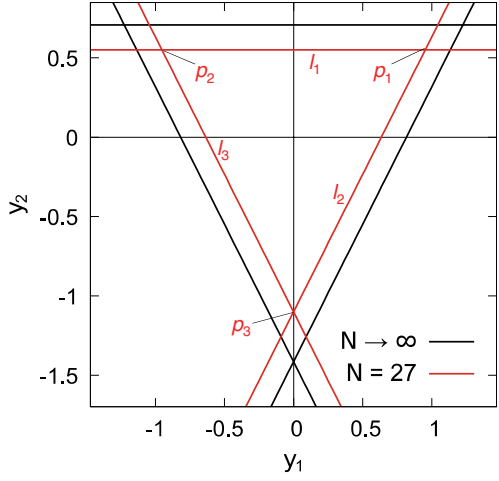}
	\caption{
	\label{fig::epsilon:yrange_3c} Possible area inside which $y_1,y_2$ can be defined.  $l_1,l_2,l_3,p_1,p_2,p_3$ denotes constituent lines and vertices at minimum system size, $N=27$. 
	}
\end{figure}

Next, we proceed to derivation for ternary system. 1-order moment of point correlation, $y_1,y_2$, should satisfy 
the conditions of 
\begin{eqnarray}
	\label{eq::epsilon:yrange_3c}
		&y_1 = \dfrac{\sqrt{6} \left(1 - x_{\rm B} - 2x_{\rm C}\right)}{2}\\[5pt]
		&y_2 = \dfrac{\sqrt{2} \left(1-3x_{\rm B}\right)}{2}\\[5pt]
		&y_2 \leq \dfrac{\sqrt{2}}{2} - \dfrac{3 \sqrt{2}}{N} := l_1\\[5pt]
		&y_2 \geq \sqrt{3} y_1 - \sqrt{2} + \dfrac{6\sqrt{2}}{N} := l_2\\[5pt]
		&y_2 \geq -\sqrt{3} y_1 - \sqrt{2} + \dfrac{6\sqrt{2}}{N} := l_3, 
\end{eqnarray}
where $y_1,y_2$ can be defined inside the area $S$ framed by three lines $l_1,l_2,l_3$, as shown in Fig.~\ref{fig::epsilon:yrange_3c}, and corresponding vertices are respectively 
\begin{eqnarray}
p_1:\left( \dfrac{\sqrt{6}}{2} - \dfrac{3\sqrt{6}}{N},~\dfrac{\sqrt{2}}{2} - \dfrac{3\sqrt{2}}{N}\right) \nonumber \\
p_2:\left( -\dfrac{\sqrt{6}}{2} + \dfrac{3\sqrt{6}}{N},~\dfrac{\sqrt{2}}{2} - \dfrac{3\sqrt{2}}{N}\right) \nonumber \\
p_3:\left(0,~-\sqrt{2}+\dfrac{6\sqrt{2}}{N}\right).
\end{eqnarray}
Inside the area $S$, we determine maximum and minimum value of $\varepsilon^2$ for pair figures with basis indices of $2_{s} \left(1,1\right),2_{s} \left(2,2\right),2_{s} \left(1,2\right)$. 
For pair figure $2_{s} \left(1,1\right)$, $\varepsilon_{11}^2$ is straightforwardly 
\begin{widetext}
\begin{eqnarray}
	\label{eq::epsilon:11_ep^2}
	\begin{split}
		&\varepsilon_{11}^2
		=
		\dfrac {D_{s}  \left( N-2 \right)  \left( N-3 \right) }{N-2 D_{s}-1}\\[5pt]
		&\hspace{20pt}
		\cdot
		\left\{
			 \left.
			 	\left( \sqrt {2}y_{2}-2 {y_{1}}^{2}+2 \right) ^{2}
			 \right/ 
			 	4\,{N}^{2}\sqrt {2}y_{2}-15\,N\sqrt {2}y_{2}+18+4\,{N}^{2}-18\,N+6\,{y_{2}}^{2}-24\,{y_{1}}^{2}
			 \right.\\
			 &\hspace{20pt}
		\underbrace{
			\left.
			 \hspace{40pt}
				+15\,\sqrt {2}y_{2}-4\,{N}^{2}\sqrt {2}{y_{1}}^{2}y_{2}+4\,{N}^{2}{y_{1}}^{4}-8\,{N}^{2}{y_{1}}^{2}+24\,{y_{1}}^{2}N+2\,{N}^{2}{y_{2}}^{2}-6\,N{y_{2}}^{2}
			\right\}
		}_{f}
	\end{split}
\end{eqnarray}
\end{widetext}
When we define denominator of $f$ in the above equation as $f_{\rm den}$, differenciation of $f$ w.r.t. $y_1$ and $y_2$ are given by 
\begin{widetext}
\begin{eqnarray}
	\label{eq::epsilon:11_f_1}
	&\begin{split}
		&\dfrac{\partial}{\partial y_1} f
		=
			\dfrac{48 \sqrt {2}\left( N-1 \right)}{f_{\rm den}^2}
			\underbrace{
				\left\{ y_{2} - \left( \dfrac{-3 \sqrt {2}+\sqrt {32 {y_{1}}^{2}+2}}{4}\right) \right\}
			}_{\rm \left(i\right)}
			\underbrace{
				\left\{ y_{2} - \left( \dfrac{-3 \sqrt {2}-\sqrt {32 {y_{1}}^{2}+2}}{4}\right) \right\}
			}_{\rm \left(ii\right)}
			\\
			&\hspace{250pt}
			\cdot
			\underbrace{
				\left\{ y_{2}- \left(\sqrt {2}{y_{1}}^{2}-\sqrt {2} \right)\right\}
			}_{\rm \left(iii\right)}
			\underbrace{
				y_{1}
			}_{\rm \left(iv\right)}:= f_1
	\end{split}&\\[5pt]
	\label{eq::epsilon:11_f_2}	
	&\begin{split}
		\dfrac{\partial }{\partial y_2} f
		=
			\dfrac{-6 \sqrt {2} \left( N-1 \right) }{f_{\rm den}^2}
			\left( 4 y_1^2 + 1 \right)
			\underbrace{
				\left\{ y_2 - {\dfrac {\sqrt {2} \left( 3 {y_{1}}^{2}-1 \right) }{4 {y_{1}}^{2}+1}} \right\}
			}_{\rm \left(v\right)}
			\underbrace{
				\left\{y_{2}- \left(\sqrt {2}{y_{1}}^{2}-\sqrt {2} \right)\right\}
			}_{\rm \left(iii\right)} := f_2.
	\end{split}
\end{eqnarray}
\end{widetext}

\begin{figure}[h]
	\centering
	\includegraphics[width=0.75\linewidth]{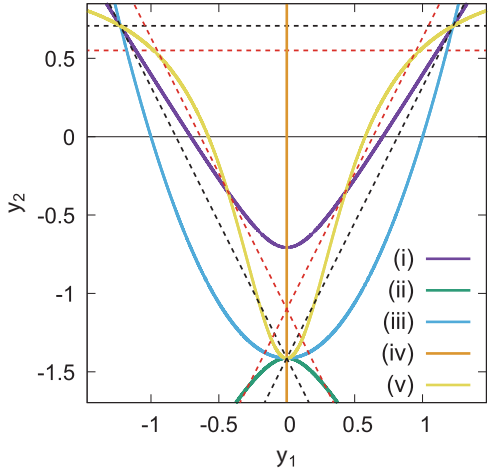}
	\caption{
		\label{fig::epsilon:yrange11}
		Relationship between $y_1$ and $y_2$ for terms (i)-(v) in Eqs.~(\ref{eq::epsilon:11_f_1}) and~(\ref{eq::epsilon:11_f_2}).
		Black and red dashed lines respectively denotes area of $S$ for $N\rightarrow\infty$ and $N=27$.
		}
\end{figure}
From Fig.~\ref{fig::epsilon:yrange11}, we can see that conditions for $f_1=f_2=0$ inside the area $S$ are restricted to the points satisfying both ${\rm \left(i\right)}=0$ and ${\rm \left(v\right)}=0$, namely,
\begin{eqnarray}
	\label{eq::epsilon11:f_1=f_2=0}
	\left(y_1,y_2\right) = \left( \pm \dfrac{\sqrt{3}}{4},-\dfrac{\sqrt{2}}{4} \right)
\end{eqnarray}
We determine whether these points provide extremum by employing the sign of Hessian for $f\left(y_1,y_2\right)$, $h\left(y_1,y_2\right)$. 
The points in Eq.~(\eqref{eq::epsilon11:f_1=f_2=0}) provides the negative value for $h$: 
\begin{eqnarray}
	\label{eq::epsilon2:h_11}
	h \left( \pm \dfrac{\sqrt{3}}{4},-\dfrac{\sqrt{2}}{4} \right)
	=
	-\dfrac {524288  \left( N-1 \right) ^{2}}{27  \left( 3 N-4 \right) ^{4} \left( N-4 \right) ^{4}}
	<
	0,	
\end{eqnarray}
which corresponds to the saddle points, i.e., they do not take extremum. 

Then, we examine extremum conditions for borders of $S$, i.e., points on lines of $l_1,l_2,l_3$.
Function $f$ for $y_2 = l_1$ is given by
\begin{widetext}
\begin{eqnarray}
	\label{eq::epsilon2:11_f_l1}
	f \left( y_2 = l_1 \right)
	=
	\dfrac { \left( 2\,{y_{1}}^{2}N-3\,N+6 \right) ^{2}}{4\,{N}^{4}{y_{1}}^{4}-12\,{N}^{4}{y_{1}}^{2}+48\,{N}^{3}{y_{1}}^{2}+9\,{N}^{4}-24\,{N}^{2}{y_{1}}^{2}-72\,{N}^{3}+198\,{N}^{2}-234\,N+108}.
\end{eqnarray}
\end{widetext}
Differentiating $f \left( y_2 = l_1 \right) $ w.r.t. $y_1$, we get 
\begin{widetext}
\begin{eqnarray}
	\label{eq::epsilon2:11_f_l1^(1)}
	\dfrac{\partial }{\partial y_1} f \left( y_2 = l_1 \right)
	=
	\dfrac{
		192\,{N}^{3} \left( N-1 \right)
		\left( {y_{1}}^{2}-\dfrac{3}{2} + \dfrac{6}{N}\right)
		\left( y_1^2 -\dfrac{3}{2} + \dfrac{15}{2N} -\dfrac{9}{N^2}\right) y_1
	}{\left\{f \left( y_2 = l_1 \right)_{\rm den}\right\}^2}.
\end{eqnarray}
\end{widetext}
Here, since we have the following conditions:
\begin{widetext}
\begin{eqnarray}
	&{y_{1}}^{2}-\dfrac{3}{2} + \dfrac{6}{N} 
	~\leq~
	\left\{\dfrac{\sqrt{6}}{2} - \dfrac{3\sqrt{6}}{N}\right\}^{2}-\dfrac{3}{2} + \dfrac{6}{N}
	~=~
	-\dfrac {6 \left(2N-9\right)}{{N}^{2}}
	~<~ 0&\\[5pt]
	&y_1^2 -\dfrac{3}{2} + \dfrac{15}{2N} -\dfrac{9}{N^2}
	~\leq~
	\left\{\dfrac{\sqrt{6}}{2} - \dfrac{3\sqrt{6}}{N}\right\}^2 -\dfrac{3}{2} + \dfrac{15}{2N} -\dfrac{9}{N^2}
	~=~
	-\dfrac {3\left(7N-30\right)}{2{N}^{2}}
	~<~ 0,
\end{eqnarray}
\end{widetext}
$f \left( y_2 = l_1 \right) $ has the following local minimum value at $y_1 = 0$:
\begin{widetext}
\begin{eqnarray}
	\label{eq::epsilon2:11_ep_l1_ex}
		\dfrac { D_s \left( N-2 \right)^2  \left( N-3 \right) }{ \left( {N}^{3}-6\,{N}^{2}+10\,N-6 \right)  \left( N-2\,D_{s}-1 \right) }
		,~
		\left(x_{\rm A},x_{\rm B},x_{\rm C}\right) = \left( \dfrac{1}{2} - \dfrac{1}{N},~\dfrac{2}{N},~\dfrac{1}{2} - \dfrac{1}{N} \right)
\end{eqnarray}
\end{widetext}

In a similar fashion, $f$ for $y_2 = l_2$ is given by
\begin{widetext}
\begin{eqnarray}
	\label{eq::epsilon2:11_f_l2}
	\begin{split}
		&f \left(y_2 = l_2\right)
		=
		\left.
		\left( Ny_{1}\,\sqrt {6}-2\,{y_{1}}^{2}N+12 \right)^2
		\right/
			4\,{N}^{4}{y_{1}}^{3}\sqrt {6}-4\,{N}^{4}{y_{1}}^{4}-6\,{N}^{4}{y_{1}}^{2}-21\,{N}^{3}y_{1}\,\sqrt {6}\\
			&\hspace{75pt}+42\,{N}^{3}{y_{1}}^{2}+69\,{N}^{2}y_{1}\,\sqrt {6}+6\,{N}^{2}{y_{1}}^{2}-72\,Ny_{1}\,\sqrt {6}-108\,{N}^{2}+396\,N-432.
	\end{split}
\end{eqnarray}
\end{widetext}
Differentiating $f \left( y_2 = l_2 \right) $ w.r.t. $y_1$, we get 
\begin{widetext}
\begin{eqnarray}
	\label{eq::epsilon2:11_f_l2^(1)}
	\begin{split}
		&\dfrac{\partial }{\partial  y_1} f \left(y_2 = l_2\right)
		=
		\dfrac{48 N^4 \left(N-1\right)}{\left\{f \left( y_2 = l_2 \right)_{\rm den}\right\}^2}
		\left(y_{1}-{\dfrac {\sqrt {6}N + \sqrt {6\,{N}^{2}+96\,N}}{4N}}\right)
		\left(y_{1}-{\dfrac {\sqrt {6}N - \sqrt {6\,{N}^{2}+96\,N}}{4N}}\right)\\
		&\hspace{125pt}
		\cdot
		\underbrace{
			\left(
				{y_{1}}^{3}-\,{\dfrac { 3 \sqrt{6} \left( {N}+24\right) {y_{1}}^{2}}{4{N}}}+\,{\dfrac { \left( 3\,{N}^{2}+48\,N-576 \right) y_{1}}{4{N}^{2}}}+{\dfrac {3\sqrt {6}}{2N}}
			\right)
		}_{g}.
	\end{split}
\end{eqnarray}
\end{widetext}
Since we have the conditions of 
\begin{widetext}
\begin{eqnarray}
	{\dfrac {\sqrt {6}N - \sqrt {6\,{N}^{2}+96\,N}}{4N}} 
	~<~
	0
	~\leq~
	y_1
	~\leq~
	\dfrac{\sqrt{6}}{2} - \dfrac{3\sqrt{6}}{N} 
	~<~ 
	{\dfrac {\sqrt {6}N + \sqrt {6\,{N}^{2}+96\,N}}{4N}}, 
\end{eqnarray}
\end{widetext}
we get
\begin{eqnarray}
	\left(y_{1}-{\dfrac {\sqrt {6}N + \sqrt {6\,{N}^{2}+96\,N}}{4N}}\right)
	\left(y_{1}-{\dfrac {\sqrt {6}N - \sqrt {6\,{N}^{2}+96\,N}}{4N}}\right)
	<
	0. \nonumber \\
	\quad
\end{eqnarray}
Since at endpoints of $g$, we have 
\begin{eqnarray}
	&g \left(y_1 = 0 \right) = \dfrac{3 \sqrt{6}}{2 N} > 0&\\[5pt]
	&g \left( y_1 = \dfrac{\sqrt{6}}{2} - \dfrac{3\sqrt{6}}{N} 	\right)
		= -\,{\dfrac {3\sqrt {6} \left( 29\,{N}^{2}-342\,N+936 \right) }{4{N}^{3
		}}} < 0, \nonumber \\
		\quad
\end{eqnarray}
it crosses with $y_1$ axes only once. 
When we explicitly consider discrete composition for $g$, we have 
\begin{eqnarray}
		&g \left(y_1 = {\dfrac {\sqrt {6} \left( N-24 \right) }{4N}} \right) = \dfrac {216\sqrt {6} \left( N-20 \right) }{{N}^{3}} > 0&\\[5pt]
		&g \left(y_1 = {\dfrac {\sqrt {6} \left( N-23 \right) }{4N}} \right) = -\dfrac {3\,\sqrt {6} \left( {N}^{2}-2192\,N+41423 \right) }{32\,{N}^{3}} < 0.\nonumber \\
		\quad 	
\end{eqnarray}
Therefore, we individually consider $\varepsilon_{11}^2 \left( y_2 = l_2 \right) $ for odd and even number of atoms, where $g$ can take minimum value at $y_1 = \left. \sqrt {6} \left( N-25 \right) \right/ 4N$, $\left. \sqrt {6} \left( N-24 \right) \right/ 4N$, $\left. \sqrt {6} \left( N-23 \right) \right/ 4N$, $\left. \sqrt {6} \left( N-22 \right) \right/ 4N$ for large $N$, respectively:
\begin{widetext}
\begin{eqnarray}
		\label{eq::epsilon2:11_ep_l2_ex1}
		&\begin{split}
			&\dfrac {  D_{s} \left( N-2 \right)  \left( N-3 \right) \left( {N}^{2}+16\,N-625 \right) ^{2}}{{N}^{2} \left( {N}^{4}+28\,{N}^{3}-1246\,{N}^{2}-13212\,N+384093 \right)  \left( N-2D_{s}-1 \right) }
			,\\
			&\hspace{50pt}\left(x_{\rm A},x_{\rm B},x_{\rm C}\right) = \left( \dfrac{1}{2} - \dfrac{21}{2N},~\dfrac{1}{2} + \dfrac{17}{2N},~\dfrac{2}{N} \right)
		\end{split}&\\[5pt]
		\label{eq::epsilon2:11_ep_l2_ex2}
		&\begin{split}
			&\dfrac { D_{s} \left( N-2 \right)  \left( N-3 \right) \left( {N}^{2}+16\,N-576 \right) ^{2}}{{N}^{2} \left( {N}^{4}+28\,{N}^{3}-1148\,{N}^{2}-12032\,N+325632 \right)  \left( N-2D_{s}-1 \right) },\\
			&\hspace{50pt}\left(x_{\rm A},x_{\rm B},x_{\rm C}\right) = \left( \dfrac{1}{2} - \dfrac{10}{N},~\dfrac{1}{2} + \dfrac{8}{N},~\dfrac{2}{N} \right)
		\end{split}&\\[5pt]
		\label{eq::epsilon2:11_ep_l2_ex3}
		&\begin{split}
			&\dfrac {  D_{s} \left( N-2 \right)  \left( N-3 \right) \left( {N}^{2}+16\,N-529 \right) ^{2}}{{N}^{2} \left( {N}^{4}+28\,{N}^{3}-1054\,{N}^{2}-10908\,N+274077 \right)  \left( N-2D_{s}-1 \right) }
			,\\
			&\hspace{50pt}\left(x_{\rm A},x_{\rm B},x_{\rm C}\right) = \left( \dfrac{1}{2} - \dfrac{19}{2N},~\dfrac{1}{2} + \dfrac{15}{2N},~\dfrac{2}{N} \right)
		\end{split}&\\[5pt]
		\label{eq::epsilon2:11_ep_l2_ex4}
		&\begin{split}
			&\dfrac { D_{s} \left( N-2 \right)  \left( N-3 \right) \left( {N}^{2}+16\,N-484 \right) ^{2}}{{N}^{2} \left( {N}^{4}+28\,{N}^{3}-964\,{N}^{2}-9840\,N+228864 \right)  \left( N-2D_{s}-1 \right) },\\
			&\hspace{50pt}\left(x_{\rm A},x_{\rm B},x_{\rm C}\right) = \left( \dfrac{1}{2} - \dfrac{9}{N},~\dfrac{1}{2} + \dfrac{7}{N},~\dfrac{2}{N} \right)
		\end{split}&
\end{eqnarray}
\end{widetext}
Final consideration is $f$ for $y_2 = l_3$, given by
\begin{widetext}
\begin{eqnarray}
	\label{eq::epsilon2:11_f_l3}
	\begin{split}
		&f \left(y_2 = l_3\right)
		=
		\left.\left( Ny_{1}\,\sqrt {6}+2\,{y_{1}}^{2}N-12 \right) ^{2} \right/ 4\,{N}^{4}{y_{1}}^{3}\sqrt {6}+6\,{N}^{4}{y_{1}}^{2}+4\,{N}^{4}{y_{1}}^{4}-42\,{N}^{3}{y_{1}}^{2}\\
		&\hspace{75pt} -21\,{N}^{3}y_{1}\,\sqrt {6}-6\,{N}^{2}{y_{1}}^{2}+69\,{N}^{2}y_{1}\,\sqrt {6}+108\,{N}^{2}-72\,Ny_{1}\,\sqrt {6}-396\,N+432.
	\end{split}
\end{eqnarray}
\end{widetext}
Differentiating $f \left( y_2 = l_3 \right) $ w.r.t. $y_1$ becomes
\begin{widetext}
\begin{eqnarray}
	\label{eq::epsilon2:11_f_l3^(1)}
	\begin{split}
		&\dfrac{\partial }{\partial y_1} f \left( y_2 = l_3 \right)
		=
		\dfrac{48 N^4 \left(N-1\right)}{\left\{f \left( y_2 = l_3 \right)_{\rm den}\right\}^2}
		\left( y_{1}+{\frac {N\sqrt {6}-\sqrt {6\,{N}^{2}+96\,N}}{4N}}\right)  \left( y_{1}+{\frac {N\sqrt {6}+\sqrt {6\,{N}^{2}+96\,N}}{4N}} \right)\\
		&\hspace{125pt}\cdot
		\underbrace{
			\left({y_{1}}^{3}+{\frac { 3\sqrt{6}\left( {N}+24\right) {y_{1}}^{2}}{4{N}}}+{\frac { \left( 3\,{N}^{2}+48\,N-576 \right) y_{1}}{4{N}^{2}}}-{\frac {3\sqrt {6}}{2N}}\right)
		}_{g}.
	\end{split}
\end{eqnarray}
\end{widetext}
Since we have the conditions of 
\begin{widetext}
\begin{eqnarray}
	-{\dfrac {\sqrt {6}N + \sqrt {6\,{N}^{2}+96\,N}}{4N}} 
	~<~
	-\dfrac{\sqrt{6}}{2} + \dfrac{3\sqrt{6}}{N} 
	~\leq~
	y_1
	~\leq~
	0
	~<~
	-{\dfrac {\sqrt {6}N - \sqrt {6\,{N}^{2}+96\,N}}{4N}},
\end{eqnarray}
\end{widetext}
we get
\begin{eqnarray}
	\left(y_{1}+{\dfrac {\sqrt {6}N + \sqrt {6\,{N}^{2}+96\,N}}{4N}}\right)
	\left(y_{1}+{\dfrac {\sqrt {6}N - \sqrt {6\,{N}^{2}+96\,N}}{4N}}\right)
	<
	0.\nonumber \\
	\quad
\end{eqnarray}
At the endpoints of $g$, we have 
\begin{eqnarray}
	&g \left( y_1 = - \dfrac{\sqrt{6}}{2} + \dfrac{3\sqrt{6}}{N} 	\right)
	= {\dfrac {3\sqrt {6} \left( 29\,{N}^{2}-342\,N+936 \right) }{4{N}^{3
	}}} > 0 \nonumber \\
	&g \left(y_1 = 0 \right) = - \dfrac{3 \sqrt{6}}{2 N} < 0,
\end{eqnarray}
which means it crosses with $y_1$ axes only once. 
Considering discrete composition for $g$ leads to 
\begin{eqnarray}
	&g \left(y_1 = -{\dfrac {\sqrt {6} \left( N-23 \right) }{4N}} \right) = \dfrac {3\,\sqrt {6} \left( {N}^{2}-2192\,N+41423 \right) }{32\,{N}^{3}} > 0 \nonumber \\
	&g \left(y_1 = -{\dfrac {\sqrt {6} \left( N-24 \right) }{4N}} \right) = -\dfrac {216\sqrt {6} \left( N-20 \right) }{{N}^{3}} < 0. 
\end{eqnarray}
Thus, in a similar fashion, considering even and odd number of atoms for $\varepsilon_{11}^2 \left( y_2 = l_3 \right) $, and it can take minimum value at $y_1 = \left. -\sqrt {6} \left( N-22 \right) \right/ 4N$, $\left. -\sqrt {6} \left( N-23 \right) \right/ 4N$, $\left. -\sqrt {6} \left( N-24 \right) \right/ 4N$, $\left. -\sqrt {6} \left( N-25 \right) \right/ 4N$ for large $N$ as follows:
\begin{widetext}
\begin{eqnarray}
	\label{eq::epsilon2:11_ep_l3_ex1}
	&\begin{split}
		&\dfrac { D_{s} \left( N-2 \right)  \left( N-3 \right) \left( {N}^{2}+16\,N-484 \right) ^{2}}{{N}^{2} \left( {N}^{4}+28\,{N}^{3}-964\,{N}^{2}-9840\,N+228864 \right)  \left( N-2D_{s}-1 \right) },\\
		&\hspace{50pt}\left(x_{\rm A},x_{\rm B},x_{\rm C}\right) = \left( \dfrac{2}{N},~\dfrac{1}{2} + \dfrac{7}{N},~\dfrac{1}{2} - \dfrac{9}{N} \right)
	\end{split}&\\[5pt]
	\label{eq::epsilon2:11_ep_l3_ex2}
	&\begin{split}
		&\dfrac {  D_{s} \left( N-2 \right)  \left( N-3 \right) \left( {N}^{2}+16\,N-529 \right) ^{2}}{{N}^{2} \left( {N}^{4}+28\,{N}^{3}-1054\,{N}^{2}-10908\,N+274077 \right)  \left( N-2D_{s}-1 \right) }
		,\\
		&\hspace{50pt}\left(x_{\rm A},x_{\rm B},x_{\rm C}\right) = \left( \dfrac{2}{N},~\dfrac{1}{2} + \dfrac{15}{2N},~\dfrac{1}{2} - \dfrac{19}{2N} \right)
	\end{split}&\\[5pt]
	\label{eq::epsilon2:11_ep_l3_ex3}
		&\begin{split}
		&\dfrac { D_{s} \left( N-2 \right)  \left( N-3 \right) \left( {N}^{2}+16\,N-576 \right) ^{2}}{{N}^{2} \left( {N}^{4}+28\,{N}^{3}-1148\,{N}^{2}-12032\,N+325632 \right)  \left( N-2D_{s}-1 \right) },\\
		&\hspace{50pt}\left(x_{\rm A},x_{\rm B},x_{\rm C}\right) = \left( \dfrac{2}{N},~\dfrac{1}{2} + \dfrac{8}{N},~\dfrac{1}{2} - \dfrac{10}{N} \right)
	\end{split}&\\[5pt]
	\label{eq::epsilon2:11_ep_l3_ex4}
	&\begin{split}
		&\dfrac {  D_{s} \left( N-2 \right)  \left( N-3 \right) \left( {N}^{2}+16\,N-625 \right) ^{2}}{{N}^{2} \left( {N}^{4}+28\,{N}^{3}-1246\,{N}^{2}-13212\,N+384093 \right)  \left( N-2D_{s}-1 \right) }
		,\\
		&\hspace{50pt}\left(x_{\rm A},x_{\rm B},x_{\rm C}\right) = \left( \dfrac{2}{N},~\dfrac{1}{2} + \dfrac{17}{2N},~\dfrac{1}{2} - \dfrac{21}{2N} \right)
	\end{split}&
	\end{eqnarray}
\end{widetext}

We finally consider $\varepsilon_{11}^2$ at vertices of $S$, $p_1,p_2,p_3$, given by
\begin{widetext}
\begin{eqnarray}
	\label{eq::epsilon2:11_ep_p1_ex}
	&\dfrac {2 D_{s}\, \left( N-2 \right)  \left( N-3 \right)  \left( 5\,N-18 \right) ^{2}}{{N}^{2} \left( N-2\,D_{s}-1 \right)  \left( 33\,{N}^{2}-277\,N+582 \right) }
	&,~
	\left(x_{\rm A},x_{\rm B},x_{\rm C}\right) = \left( 1 - \dfrac{4}{N},~\dfrac{2}{N},~\dfrac{2}{N} \right)\\[5pt]
	\label{eq::epsilon2:11_ep_p2_ex}
	&\dfrac {4 D_{s}\, \left( N-2 \right)  \left( N-3 \right)}{\left( N-2\,D_{s}-1 \right)  \left( 3\,{N}^{2}-11\,N+12 \right) }
	&,~
	\left(x_{\rm A},x_{\rm B},x_{\rm C}\right) = \left(\dfrac{2}{N},~ 1 - \dfrac{4}{N},~\dfrac{2}{N} \right)\\[5pt]
	\label{eq::epsilon2:11_ep_p3_ex}
	&\dfrac {2 D_{s}\, \left( N-2 \right)  \left( N-3 \right)  \left( 5\,N-18 \right) ^{2}}{{N}^{2} \left( N-2\,D_{s}-1 \right)  \left( 33\,{N}^{2}-277\,N+582 \right) }
	&,~
	\left(x_{\rm A},x_{\rm B},x_{\rm C}\right) = \left(\dfrac{2}{N},~\dfrac{2}{N},~ 1 - \dfrac{4}{N} \right)
\end{eqnarray}
\end{widetext}
From above Eqs.~\eqref{eq::epsilon2:11_ep_l1_ex}, \eqref{eq::epsilon2:11_ep_l2_ex1}, \eqref{eq::epsilon2:11_ep_l2_ex2}, \eqref{eq::epsilon2:11_ep_l2_ex3}, \eqref{eq::epsilon2:11_ep_l2_ex4}, \eqref{eq::epsilon2:11_ep_l3_ex1}, \eqref{eq::epsilon2:11_ep_l3_ex2}, \eqref{eq::epsilon2:11_ep_l3_ex3}, \eqref{eq::epsilon2:11_ep_l3_ex4}, \eqref{eq::epsilon2:11_ep_p1_ex}, \eqref{eq::epsilon2:11_ep_p2_ex} and \eqref{eq::epsilon2:11_ep_p3_ex}, maximum and minimum values for $\epsilon^2$ are given by
\begin{widetext}
\begin{eqnarray}
	&{\rm min}\left( \varepsilon_{11}^2 \right) \sim \dfrac{D_s}{N},~\left(x_{\rm A},x_{\rm B},x_{\rm C}\right) \sim 
	\left(0,~\dfrac{1}{2},~\dfrac{1}{2} \right),
	\left(~\dfrac{1}{2},~0,~\dfrac{1}{2} \right),
	\left(\dfrac{1}{2},~\dfrac{1}{2},~0 \right)&\\[5pt]
	&{\rm max}\left( \varepsilon_{11}^2 \right) \sim \dfrac{50 D_s}{33N},~\left(x_{\rm A},x_{\rm B},x_{\rm C}\right) \sim 
	\left(1,~0,~0 \right),
	\left(0,~0,~1 \right).
\end{eqnarray}
\end{widetext}

\subsubsection{Basis index set of $\left(22\right)$}
$\varepsilon_{22}^2$ for a certain pair figure $2_s \left(2,2\right)$ is given by
\begin{widetext}
\begin{eqnarray}
	\label{eq::epsilon:22_ep^2}
	\varepsilon_{22}^2
	=
	{\dfrac {2D_{s}\, \left( N-2 \right)  \left( N-3 \right) }{N-2\,D_{s}-1}}
	\cdot
	\underbrace{
		\dfrac { \left( -2\,y_{2}+\sqrt {2} \right)  \left( y_{2}+\sqrt {2}\right) }{ \left( \sqrt {2}N-2\,Ny_{2}-3\,\sqrt {2} \right)  \left( 2\,\sqrt {2}N+2\,Ny_{2}-3\,\sqrt {2} \right) }
	}_{f}.
\end{eqnarray}
\end{widetext}
Differentiating $f$ w.r.t. $y_2$ leads to 
\begin{eqnarray}
	\label{eq::epsilon:22_f_1}
	\dfrac{\partial }{\partial y_2} f 
	=
	\dfrac { 72 \left(N-1\right)  \left( y_{2}+\dfrac{\sqrt {2}}{4} \right) }{ \left( \sqrt {2}N-2\,Ny_{2}-3\,\sqrt {2} \right) ^{2} \left( 2\,\sqrt {2}N+2\,Ny_{2}-3\,\sqrt {2} \right) ^{2}},\nonumber \\
	\quad
\end{eqnarray}
which means it takes local minimum at $y_2 = -\left.\sqrt{2}\right/4$:
\begin{eqnarray}
	\label{eq::epsilon2:22_ex1}
	\dfrac{D_s \left( N-3 \right)}{\left( N-2 \right) \left( N-2 D_s - 1\right)},~x_{\rm B} = \dfrac{1}{2}.
\end{eqnarray}
At the endpoints of defined $y_2$, $\varepsilon_{22}^2$ takes 
\begin{eqnarray}
	\label{eq::epsilon2:22_p1}
	&\dfrac{2 D_s \left( N-2 \right)^2}{N^2 \left( N-2 D_s - 1\right)}, x_{\rm B} = \dfrac{2}{N}&\\[5pt]
	\label{eq::epsilon2:22_p2}
	&\dfrac{4 D_s \left( N-2 \right) \left( N-3 \right) \left( N-4 \right)}{3 N^2 \left( N-5 \right) \left( N-2 D_s - 1\right)}, x_{\rm B} = \dfrac{N-4}{N}.
\end{eqnarray}
From Eqs.~\eqref{eq::epsilon2:22_ex1}, \eqref{eq::epsilon2:22_p1} and \eqref{eq::epsilon2:22_p2}, 
maximum and minimum values for $\varepsilon_{22}^2$ are given by
\begin{eqnarray}
	&{\rm min}\left( \varepsilon_{22}^2 \right) \sim \dfrac{D_s}{N},~x_{\rm B} = \dfrac{1}{2}&\\[5pt]
	&{\rm max}\left( \varepsilon_{22}^2 \right) \sim \dfrac{2 D_s}{N},~x_{\rm B} \sim 0.
\end{eqnarray}

\subsubsection{Basis index set of $\left(12\right)$}
$\varepsilon_{12}^2$ is given by 
\begin{widetext}
\begin{eqnarray}
	\label{eq::epsilon:12_ep^2}
	\begin{split}
		&\varepsilon_{12}^2 = {\dfrac {4 D_{s}\, \left( N-2 \right)  \left( N-3 \right) }{N-2\,D_{s}-1}}\\
		&\hspace{20pt}\cdot
	 	\left. {y_{1}}^{2} \left( y_{2}-\dfrac{\sqrt {2}}{2} \right)
	 	\right/
	 	\left(
	 	{N}^{2}\sqrt {2}{y_{1}}^{2}-\sqrt {2}{N}^{2}{y_{2}}^{2}+4\,{N}^{2}{y_{1}}^{2}y_{2}-3\sqrt {2}N{y_{1}}^{2}\,
	 	\right.\\
	 	&\hspace{20pt}
	 	\underbrace{\left.
	 	\hspace{45pt}+\sqrt {2}N{y_{2}}^{2}-2\,{N}^{2}\sqrt {2}-4\,{N}^{2}y_{2}+8\,\sqrt {2}N+10\,Ny_{2}-6\,\sqrt {2}-6\,y_{2}\right)
	 	}_{f}
	\end{split}
\end{eqnarray}
\end{widetext}
Differentiating $f$ w.r.t. $y_1$ and $y_2$ leads to 
\begin{widetext}
\begin{eqnarray}
	\label{eq::epsilon:12_f_1}
	&\dfrac{\partial }{\partial y_1} f
	=
	\dfrac{
		-2\,\sqrt {2}N \left( N-1 \right)  \left( y_{2}+\sqrt {2}-{\dfrac {3\sqrt {2}}{N}} \right)  \left( y_{2}+\sqrt {2} \right)  \left( y_{2}-\dfrac{\sqrt {2}}{2} \right) y_{1}
	}{f_{\rm den}^2} := f_1&\\[10pt]
	\label{eq::epsilon:12_f_2}
	&\begin{split}
		&\dfrac{\partial }{\partial y_2} f
		=\dfrac{\sqrt {2}N \left( N-1 \right)}{f_{\rm den}^2}\\
		&\hspace{25pt}\cdot
		\left\{  \left( y_{2}-\,{\dfrac {\sqrt {2}N+3\,\sqrt {2\,{N}^{2}-4\,N}}{2N}} \right)  \left( y_{2}-\,{\dfrac {\sqrt {2}N-3\,\sqrt {2\,{N}^{2}-4\,N}}{2N}} \right) +3\,{y_{1}}^{2} \right\} {y_{1}}^{2} :=f_2. 
	\end{split}
\end{eqnarray}
\end{widetext}
In Eq.~\eqref{eq::epsilon:12_f_1}, $f_1=0$ is satisfied only when $y_1=0$, and we have the conditions of 
\begin{widetext}
\begin{eqnarray}
	&\dfrac {\sqrt {2}N-3\,\sqrt {2\,{N}^{2}-4\,N}}{2N}
	~<~
	\dfrac {\sqrt {2} N - 3\,\sqrt {2\,\left(N-4\right)^2}}{2N}
	~=~
	- \sqrt{2} + \dfrac{6\sqrt{2}}{N}
	~\leq~ y_2
	&\\
	&\dfrac {\sqrt {2}N+3\,\sqrt {2\,{N}^{2}-4\,N}}{2N}
	~>~
	\dfrac{\sqrt{2}}{2}
	~\geq~
	y_2,
\end{eqnarray}
\end{widetext}
both $f_1=0$ and $f_2=0$ are satisfied only when $y_1=0$, whose Hessian takes
\begin{eqnarray}
	h \left(y_1 = 0\right) = 0
\end{eqnarray}
that cannot determine whether it corresponds to extremum. 
Meanwhile, since we have $f \left(y_1 = 0\right) = 0$, it is not extremum. 
Therefore, there is no extremum inside the area of $S$. 

Next, we examine extremum condition for borders of $l_1,l_2,l_3$. 
$f$ for $y_2 = l_1$ is given by
\begin{widetext}
\begin{eqnarray}
	f \left(y_2 = l_1\right) = {-\dfrac {2 y_1^2}{  2\,{N}^{3}{y_{1}}^{2}-10\,{N}^{2}{y_{1}}^{2}-3\,{N}^{3}+21\,{N}^{2}-42\,N+24 }}.
\end{eqnarray}
\end{widetext}
Differentiating $f \left(y_2 = l_1\right)$ w.r.t. $y_1$ leads to
\begin{widetext}
\begin{eqnarray}
	\dfrac{\partial }{\partial y_1} f \left(y_2 = l_1\right) = {\dfrac {12 \left( N-1 \right)  \left( N-2 \right)  \left( N-4\right) y_{1}}{ \left( 2\,{N}^{3}{y_{1}}^{2}-10\,{N}^{2}{y_{1}}^{2}-3\,{N}^{3}+21\,{N}^{2}-42\,N+24 \right) ^{2}}}.
\end{eqnarray}
\end{widetext}
Therefore, $\varepsilon_{12}^2 \left(y_2 = l_1\right)$ takes minimum value at $y_1 = 0$: 
\begin{eqnarray}
	\label{eq::epsilon2:12_ep_l1_ex}
	0
	,~
	\left(x_{\rm A},x_{\rm B},x_{\rm C}\right) = \left(\dfrac{1}{2}-\dfrac{1}{N},~\dfrac{2}{N},~ \dfrac{1}{2}-\dfrac{1}{N} \right).
\end{eqnarray}
Then, $f$ for $y_2 = l_2$ is given by 
\begin{widetext}
\begin{eqnarray}
	f \left(y_2 = l_2\right) = {\dfrac { \left( \sqrt {6}N-2\,Ny_{1}-4\,\sqrt {6} \right) {y_{1}}^{2}}{4\,{N}^{3}\sqrt {6}{y_{1}}^{2}-8\,{N}^{3}{y_{1}}^{3}-16\,\sqrt {6}{y_{1}}^{2}{N}^{2}+36\,{N}^{2}y_{1}+24\,N\sqrt {6}-36\,Ny_{1}-24\,\sqrt {6}}}.
\end{eqnarray}
\end{widetext}
Differentiating $f \left(y_2 = l_2\right)$ w.r.t. $y_1$ leads to
\begin{widetext}
\begin{eqnarray}
	\begin{split}
		&\dfrac{\partial }{\partial y_1} f \left(y_2 = l_2\right)
		=
			{\dfrac {-9{N}^{2} \left( N-1 \right) y_{1}}{ \left( {N}^{3}\sqrt {6}{y_{1}}^{2}-2\,{N}^{3}{y_{1}}^{3}-4\,\sqrt {6}{y_{1}}^{2}{N}^{2}+9\,{N}^{2}y_{1}+6\,N\sqrt {6}-9\,Ny_{1}-6\,\sqrt {6} \right) ^{2}}}\\
		&\hspace{220pt} \cdot 
		\underbrace{
			\left({y_{1}}^{2}-{\frac { \left( {N}^{2}\sqrt {6}-8\,N\sqrt {6}\right) y_{1}}{4{N}^{2}}}-{\frac {2\,N-8}{{N}^{2}}}\right)
		}_{g}.
	\end{split}
\end{eqnarray}
\end{widetext}
At the endpoints for $g$, we have 
\begin{eqnarray}
	&g \left( y_1 = 0 \right)
	= \dfrac{-2 N +8}{N^2} < 0 &\\[5pt]
	&g \left(y_1 = \dfrac{\sqrt{6}}{2} - \dfrac{3\sqrt{6}}{N} \right) = \dfrac{\left( N-4 \right) \left( 3 N - 26 \right)}{4 N^2} > 0, 
\end{eqnarray}
which means it crosses with $y_1$ axes only once.
Considering discrete compositions for $g$, we get 
\begin{eqnarray}
	&g \left( y_1 = \dfrac{\sqrt{6} \left( N-3 \right)}{4N} \right) = \dfrac{- N + 19}{8 N^2} < 0&\\[5pt]
	&g \left( y_1 = \dfrac{\sqrt{6} \left( N-2 \right)}{4N} \right) = \dfrac{ N + 14 }{4 N^2} > 0.
\end{eqnarray}
Thus, we consider $\varepsilon_{12}^2 \left(y_2 = l_2\right)$ for even and odd number of atoms respectively, which can take maximum value at $y_1 = \left. \sqrt{6} \left(N-4\right) \right/ 4N$, $\left. \sqrt{6} \left(N-2\right) \right/ 4N$, $\left. \sqrt{6} \left(N-3\right) \right/ 4N$, $\left. \sqrt{6} \left(N-1\right) \right/ 4N$ for large $N$:
\begin{widetext}
\begin{eqnarray}
	\label{eq::epsilon2:12_ep_l2_ex2}
		&\dfrac {D_{s}\, \left( N-3 \right) \left( N-4 \right) ^{3}}{ \left( N-2\,D_{s}-1 \right)  \left( {N}^{2}+2\,N+24 \right) {N}^{2}}
		&,~
		\left(x_{\rm A},x_{\rm B},x_{\rm C}\right) = \left(\dfrac{1}{2},~\dfrac{1}{2}-\dfrac{2}{N},~\dfrac{2}{N} \right)\\[5pt]
	\label{eq::epsilon2:12_ep_l2_ex3}
		&\dfrac {D_{s}\, \left( N-5 \right)  \left( N-3 \right)  \left( N-2 \right) ^{3}}{ \left( N-2\,D_{s}-1 \right)  \left( {N}^{3}+{N}^{2}+23\,N-41 \right) {N}^{2}}
		&,~
		\left(x_{\rm A},x_{\rm B},x_{\rm C}\right) = \left(\dfrac{1}{2}+\dfrac{1}{2N},~\dfrac{1}{2}-\dfrac{5}{2N},~\dfrac{2}{N} \right)\\[5pt]
	\label{eq::epsilon2:12_ep_l2_ex1}
		&\dfrac {D_{s}\, \left( N-6 \right)  \left( N-3 \right)  \left( N-2\right) ^{3}}{ \left( N-2\,D_{s}-1 \right)  \left( {N}^{3}+2\,{N}^{2}+24\,N-32 \right) {N}^{2}}
		&,~
		\left(x_{\rm A},x_{\rm B},x_{\rm C}\right) = \left(\dfrac{1}{2}+\dfrac{1}{N},~\dfrac{1}{2}-\dfrac{3}{N},~\dfrac{2}{N} \right)\\[5pt]
	\label{eq::epsilon2:12_ep_l2_ex4}
		&\dfrac {D_{s}\, \left( N-7 \right)  \left( N-3 \right)  \left( N-2\right) \left( N-1\right)}{ \left( N-2\,D_{s}-1 \right)  \left( {N}^{2}+4N +27 \right) {N}^{2}}
		&,~
		\left(x_{\rm A},x_{\rm B},x_{\rm C}\right) = \left(\dfrac{1}{2}+\dfrac{3}{2N},~\dfrac{1}{2}-\dfrac{7}{2N},~\dfrac{2}{N} \right).
\end{eqnarray}
\end{widetext}
Finally, $f$ for $y_2 = l_3$ becomes
\begin{widetext}
\begin{eqnarray}
	f \left(y_2 = l_3\right)
	=
	{\dfrac { \left( \sqrt {6}N+2\,Ny_{1}-4\,\sqrt {6} \right) {y_{1}}^{2}}{4\,{N}^{3}\sqrt {6}{y_{1}}^{2}+8\,{N}^{3}{y_{1}}^{3}-16\,\sqrt {6}{y_{1}}^{2}{N}^{2}-36\,{N}^{2}y_{1}+24\,N\sqrt {6}+36\,Ny_{1}-24\,\sqrt {6}}}.
\end{eqnarray}
\end{widetext}
Differentiating $f \left(y_2 = l_3\right)$ w.r.t. $y_1$ leads to 
\begin{widetext}
\begin{eqnarray}
	\begin{split}
	&\dfrac{\partial }{\partial y_1} f \left(y_2 = l_3\right)
	=
		{\dfrac {-9{N}^{2} \left( N-1 \right) y_{1}}
		{ \left( {N}^{3}\sqrt {6}{y_{1}}^{2}+2\,{N}^{3}{y_{1}}^{3}-4\,\sqrt {6}{y_{1}}^{2}{N}^{2}-9\,{N}^{2}y_{1}+6\,N\sqrt {6}+9\,Ny_{1}-6\,\sqrt {6} \right) ^{2}}}\\
	&\hspace{220pt} \cdot 
		\underbrace{
			\left({y_{1}}^{2}+{\frac { \left( {N}^{2}\sqrt {6}-8\,N\sqrt {6}\right) y_{1}}{4{N}^{2}}}-{\frac {2\,N-8}{{N}^{2}}}\right)
		}_{g}.
	\end{split}
\end{eqnarray}
\end{widetext}
At the endpoints of $g$, we have
\begin{eqnarray}
	&g \left(y_1 = - \dfrac{\sqrt{6}}{2} + \dfrac{3\sqrt{6}}{N} \right) = -\dfrac{\left( N-4 \right) \left( 3 N - 26 \right)}{4 N^2} > 0&\nonumber \\
	&g \left( y_1 = 0 \right)
	= \dfrac{- 2 N + 8}{N^2} < 0,
\end{eqnarray}
which means it crosses with $y_1$ axes only once. 
Considering discrete composition fo $g$, we get 
\begin{eqnarray}
	&g \left( y_1 = -\dfrac{\sqrt{6} \left( N-2 \right)}{4N} \right) = \dfrac{ N + 14 }{4 N^2} > 0&\\[5pt]
	&g \left( y_1 = -\dfrac{\sqrt{6} \left( N-3 \right)}{4N} \right) = \dfrac{- N + 19}{8 N^2} < 0.
\end{eqnarray}
Thus, in a similar fashion, $\varepsilon_{12}^2 \left(y_2 = l_3\right)$ can take maximum value at $y_1 = \left. -\sqrt{6} \left(N-1\right) \right/ 4N$, $\left. -\sqrt{6} \left(N-2\right) \right/ 4N$, $y_1 = \left. -\sqrt{6} \left(N-3\right) \right/ 4N$, $\left. -\sqrt{6} \left(N-4\right) \right/ 4N$ for large $N$: 
\begin{widetext}
\begin{eqnarray}
	\label{eq::epsilon2:12_ep_l3_ex1}
		&\dfrac {D_{s}\, \left( N-7 \right)  \left( N-3 \right)  \left( N-2\right) \left( N-1\right) }{ \left( N-2\,D_{s}-1 \right)  \left( {N}^{2}+4\,N+27 \right) {N}^{2}}
		&,~
		\left(x_{\rm A},x_{\rm B},x_{\rm C}\right) = \left(\dfrac{2}{N},~\dfrac{1}{2}-\dfrac{7}{2N},~\dfrac{1}{2}+\dfrac{3}{2N} \right)\\[5pt]
	\label{eq::epsilon2:12_ep_l3_ex2}
		&\dfrac {D_{s}\, \left( N-6 \right)  \left( N-3 \right)  \left( N-2\right) ^{3}}{ \left( N-2\,D_{s}-1 \right)  \left( {N}^{3}+2\,{N}^{2}+24\,N-32 \right) {N}^{2}}
		&,~
		\left(x_{\rm A},x_{\rm B},x_{\rm C}\right) = \left(\dfrac{2}{N},~\dfrac{1}{2}-\dfrac{3}{N},~\dfrac{1}{2}+\dfrac{1}{N} \right)\\[5pt]
	\label{eq::epsilon2:12_ep_l3_ex3}
		&\dfrac {D_{s}\, \left( N-5 \right)  \left( N-2 \right)  \left( N-3\right) ^{3}}{ \left( N-2\,D_{s}-1 \right)  \left( {N}^{3}+{N}^{2}+23\,N-41 \right) {N}^{2}}
		&,~
		\left(x_{\rm A},x_{\rm B},x_{\rm C}\right) = \left(\dfrac{2}{N},~\dfrac{1}{2}-\dfrac{5}{2N},~\dfrac{1}{2}+\dfrac{1}{2N} \right)\\[5pt]
	\label{eq::epsilon2:12_ep_l3_ex4}
		&\dfrac {D_{s}\, \left( N-3 \right)  \left( N-4\right) ^{3}}{ \left( N-2\,D_{s}-1 \right)  \left( {N}^{2}+2\,N+24 \right) {N}^{2}}
		&,~
		\left(x_{\rm A},x_{\rm B},x_{\rm C}\right) = \left(\dfrac{2}{N},~\dfrac{1}{2}-\dfrac{2}{N},~\dfrac{1}{2} \right)
\end{eqnarray}
\end{widetext}
Finally, at vertices of $S$, $p_1,p_2,p_3$, $\varepsilon_{12}^2$ is given by 
\begin{widetext}
\begin{eqnarray}
	\label{eq::epsilon2:12_ep_p1_ex}
		&{\dfrac { 2 D_{s} \left( N-2 \right) \left( N-3 \right) \left( N-6 \right) ^{2} }{{N}^{2} \left( 5\,{N}^{2}-41\,N+86 \right) \left( N-2\,D_{s}-1 \right) }}
		&,~
		\left(x_{\rm A},x_{\rm B},x_{\rm C}\right) = \left( 1 - \dfrac{4}{N},~\dfrac{2}{N},~\dfrac{2}{N} \right)\\[5pt]
	\label{eq::epsilon2:12_ep_p2_ex}
		&0
		&,~
		\left(x_{\rm A},x_{\rm B},x_{\rm C}\right) = \left(\dfrac{2}{N},~ 1 - \dfrac{4}{N},~\dfrac{2}{N} \right)\\[5pt]
	\label{eq::epsilon2:12_ep_p3_ex}
		&{\dfrac { 2 D_{s} \left( N-2 \right) \left( N-3 \right) \left( N-6 \right) ^{2} }{{N}^{2} \left( 5\,{N}^{2}-41\,N+86 \right) \left( N-2\,D_{s}-1 \right) }}
		&,~
		\left(x_{\rm A},x_{\rm B},x_{\rm C}\right) = \left(\dfrac{2}{N},~\dfrac{2}{N},~ 1 - \dfrac{4}{N} \right).
\end{eqnarray}
\end{widetext}
From above Eqs.~\eqref{eq::epsilon2:12_ep_l1_ex}, \eqref{eq::epsilon2:12_ep_l2_ex1}, \eqref{eq::epsilon2:12_ep_l2_ex2}, \eqref{eq::epsilon2:12_ep_l2_ex3}, \eqref{eq::epsilon2:12_ep_l2_ex4}, \eqref{eq::epsilon2:12_ep_l3_ex1}, \eqref{eq::epsilon2:12_ep_l3_ex2}, \eqref{eq::epsilon2:12_ep_l3_ex3}, \eqref{eq::epsilon2:12_ep_l3_ex4}, \eqref{eq::epsilon2:12_ep_p1_ex}, \eqref{eq::epsilon2:12_ep_p2_ex} and \eqref{eq::epsilon2:12_ep_p3_ex}, maximum and minimum values for $\varepsilon_{12}^2$ are given by 
\begin{widetext}
\begin{eqnarray}
	&{\rm min}\left( \varepsilon_{12}^2 \right) = 0,~x_{\rm A} = x_{\rm C}&\\[5pt]
	&{\rm max}\left( \varepsilon_{12}^2 \right) \sim \dfrac{D_s}{N},~\left(x_{\rm A},x_{\rm B},x_{\rm C}\right) \sim 
	\left( \dfrac{1}{2},~\dfrac{1}{2},~0 \right),
	\left( 0,~\dfrac{1}{2},~\dfrac{1}{2} \right).
\end{eqnarray}
\end{widetext}

\end{document}